\documentclass[11pt, oneside]{article}   	
\usepackage{geometry}                		
\usepackage{graphicx}				
\usepackage{amssymb}
\usepackage{url}
\usepackage{hyperref}
\usepackage{tikz}
\usepackage{graphicx}
\usepackage{caption}
\usepackage{amsmath}
\usepackage{algorithm}
\usepackage{algpseudocode}
\usepackage{mathtools}
\usepackage{authblk}

\newtheorem{mydef}{Definition}
\newtheorem{mylem}{Lemma}
\newtheorem{mycor}{Corollary}
\newtheorem{mytheo}{Theorem}
\newenvironment{myproof}{\paragraph{Proof:}}{\hfill$\square$}
\newcommand{\pushright}[1]{\ifmeasuring@#1\else\omit\hfill$\displaystyle#1$\fi\ignorespaces}
\newcommand\myeq{\stackrel{\mathclap{\normalfont\mbox{def}}}{=}}

\title{Overlapping Community Detection via Local Spectral Clustering}
\author[1]{Yixuan Li\thanks{yli@cs.cornell.edu}}
\author[2]{Kun He\thanks{brooklet60@hust.edu.cn}}
\author[1]{David Bindel\thanks{bindel@cs.cornell.edu}}
\author[1]{John Hopcroft\thanks{jeh@cs.cornell.edu}}
\affil[1]{Cornell University}
\affil[2]{ Huazhong Science and Technology University}

\begin{document}
\maketitle

\begin{abstract}
Large graphs arise in a number of contexts and understanding their structure and extracting information from them is an important research area. Early algorithms on mining communities have focused on the global structure, and often run in time functional to the size of the entire graph. Nowadays, as we often explore networks with billions of vertices and find communities of size hundreds, it is crucial to shift our attention from macroscopic structure to microscopic structure in large networks. A growing body of work has been adopting local expansion methods in order to identify the community members from a few exemplary seed members.

In this paper, we propose a novel approach for finding overlapping communities called \textsc{Lemon} (Local Expansion via Minimum One Norm).
The algorithm finds the community by seeking a sparse vector in the span of the local spectra such that the seeds are in its support.
We show that \textsc{Lemon} can achieve the highest detection accuracy among state-of-the-art proposals. The running time depends on the size of the community rather than that of the entire graph. The algorithm is easy to implement, and is highly parallelizable. We further provide theoretical analysis on the local spectral properties, bounding the measure of tightness of extracted community in terms of the eigenvalues of graph Laplacian. 

Moreover, given that networks are not all similar in nature, a comprehensive analysis on how the local expansion approach is suited for uncovering communities in different networks is still lacking. We thoroughly evaluate our approach using both synthetic and real-world datasets across different domains, and analyze the empirical variations when applying our method to inherently different networks in practice.  In addition, the heuristics on how the seed set quality and quantity would affect the performance are provided.
\end{abstract}

\section{Introduction}

Analyzing the structure and extracting information from complex networks is an important research area. Significant research has been carried out in finding the structure of networks and identifying communities \cite{fortunato2010community}.

In early work, researchers assumed that communities were disjoint and had more internal connections than external connections. Both assumptions have been discarded since it is clear that in most networks a vertex belongs to more than one community. For instance, in social networks, one might belong to a work community, a community of friends, and a community of individuals that share the same hobby such as golf; in co-purchased networks, one item might belong to multiple categories. Also since we are dealing with networks with hundreds of millions of vertices,  an individual in a community of size 100\footnote{A statistical study on social networks done by Leskovec \emph{et al.} \cite{leskovec2008statistical} has shown that  real-world communities with high quality are quite small and usually consist of no more than 100 vertices.} will certainly have more links outside the community than inside. These key insights have motivated us to identify communities from a new perspective.

Considerable researches on detecting communities have focused on the global structure. And these globally based detection algorithms usually run in time functional to the size of the entire graph, a major drawback in computational cost. Nowadays, we explore networks with billions of vertices to find communities of size a hundred. Thus, taking the entire graph into account might not serve as a practical solution in many situations. It is thus crucial to shift our attention from global structure to local structure in large networks, and develop new approaches that enable finding communities in time functional to the size of the community.

Quite recently, there has been a growing interest in finding communities by locally expanding an exemplary seed set in the community of interest \cite{andersen2006communities}\cite{Isabel:KDD14}\cite{whang2013overlapping}\cite{Jaewon:ICDM2012}.  This type of algorithm usually starts with a few members that are already known to be in the target community, and the goal is to uncover the remaining members in the community as the exemplary members. These known members are usually referred to as \emph{seeds} in the literature, and the process of growing the seed set gradually into a larger set until the target community is revealed is called \emph{seed set expansion}. The setting of seed set expansion can be widely applied to real world applications. For example, in web search, with a few known pages that share similar information, we could generate a larger group of web pages that contains the relevant contents with respect to a certain search query; in product networks, seed set expansion enables the automatical categorizing of products that are discovered to be in the same community as the labeled items.

The random walk technique has been extensively adopted as a subroutine for locally growing the seed set in the literature \cite{andersen2006communities}\cite{jin2011markov}\cite{Isabel:KDD14}\cite{pons2005computing}\cite{rosvall2011multilevel}\cite{whang2013overlapping}\cite{Jaewon:ICDM2012}.  The dynamics of random walks are effective in finding a local community since they make non-uniform expansion decisions based on the structure revealed during the exploration of the neighborhood surrounding the seeds \cite{andersen2006communities}. This implies that random walk based local expansion is able to trace the  community members in a principled way that best resembles the natural process for forming the local community structure. Very recently, Abrahao \emph{et al.}    also experimentally verified that random walk produces communities that are most structurally similar to real-world communities amongst various algorithmic communities  \cite{abrahao2014separability}.

In this paper we propose a novel approach for finding overlapping communities called {\textsc {Lemon}} (Local Expansion via Minimum One Norm)\footnote{Our demo code is publicly available at: \url{https://github. com/yixuanli/lemon}.} for finding overlapping communities in large networks. We systematically  demonstrate that \textsc{Lemon} can achieve both high efficiency and effectiveness that significantly stands out amongst state-of-the-art proposals. Specifically, we consider the span of a few dimensions of vectors after the short random walk and use it as the approximate invariant subspace, which we refer as \emph{local spectra}.

In contrast to the traditional spectral clustering methods, our local spectral method does not require the burdensome computation of a large number of singular vectors. In addition, as traditional spectral methods usually partition the vertices into disjoint communities, we make another fundamental change. Concretely, we mine the communities from the subspace by seeking a sparse approximate indicator vector in the span of the local spectral such that the seeds are in its support. In practice, this can be mathematically achieved by solving a $\ell^1$-penalized linear programming problem.

We aim to develop a comprehensive understanding of the local spectral approach for identifying a community from a small seed set. Following the central idea of our approach, we seek to solve fundamentally important questions such as:  what defines ``good" communities and when do they emerge as we expand the seed set (Section \ref{stop})? How to find a small community in time functional to the size of the community rather than  that of the entire graph (Section \ref{sample})? What defines ``good" seeds and how many seeds could uniquely define a community (Section \ref{seed})?

And given that networks are not all similar in nature, how the local expansion approach is suited for uncovering communities in different types of networks (Section \ref{empirical_comparison})?

We thoroughly evaluate our approach using both synthetic and real-world datasets across different domains, and analyze the empirical variations when applying our method to inherently different networks in practice. We believe that the insights we gained from researching on these problems would provide valuable guidance for future investigation on this topic.

\section{Related Work}

A considerable amount of literature has been published on finding communities in large social and information networks. We highlight a few ideas that have recently emerged in the literature to clarify how our method differs.

\textbf{Globally based community finding algorithms.} Various community detection algorithms have been developed in the past decade. And most of the algorithms fall into the category of global approach. One stream of global algorithms attempt to find communities by optimizing an objective function. For example, GCE \cite{lee2010detecting} identifies maximal cliques as seed communities. It expands these cliques by greedily optimizing a local fitness function. OSLOM \cite{lancichinetti2011finding} is also based on the optimization of a fitness function, which  expresses the statistical significance of clusters with respect to random fluctuations (i.e., the random graph generated by the configuration model \cite{molloy1995critical} during community expansion).  However, the communities identified by mathematical construction may structurally diverge from real communities as pointed in \cite{abrahao2014separability}. Another main stream of research adopts the label propagation approach \cite{raghavan2007near}, which defines rules that simulate the spread of labels of vertices in the network. The DEMON  algorithm  \cite{coscia2012demon}, for example, democratically lets each vertex vote for the communities it sees surrounding it in its limited view of the global system using a label propagation algorithm, and then merges the local communities into a global collection. Other approaches such as Link Community (LC) \cite{ahn2010link} partitions the graph by first building a hierarchical  link dendrogram according to the link similarity and then cutting the dendrogram at some threshold to yield link communities. 

\textbf{Random walk based detection algorithms.} As noted in the preceding section, among the divergent approaches, random walks tend to reveal communities that bear the closest resemblance to the ground truth communities in nature \cite{abrahao2014separability}.  In the following, we briefly review some methods that have adopted the random walk technique in finding communities. Speaking of methods that focus on the global structure, Pons \emph{et al.} \cite{pons2005computing} proposed a hierarchical agglomerative algorithm, \emph{WalkTrap}, that quantified the similarity between vertices using random walks and then partitioned the network into non-overlapping communities. Meil$\check{a}$ \emph{et al.}  \cite{meila2001random} presented a  clustering approach by viewing the pairwise similarities as edge flows in a random walk and studied the eigenvectors and values of the resulting transition matrix.   A later successful algorithm, \emph{Infomap}, proposed by by Rosvall \& Bergstrom \cite{rosvall2011multilevel} enables uncovering hierarchical structures in networks by compressing a description of a random walker as a proxy for real flow on networks. Variants of this technique such as biased random walk \cite{zlatic2010topologically} has also been employed in community finding.

 \textbf{Local expansion based approaches.} To interpret the problem of community detection from a local perspective, our work shares the same spirit as the local expansion algorithms in \cite{andersen2006communities}, \cite{Isabel:KDD14}, \cite{whang2013overlapping}
and \cite{kloster2014heat}.
Specifically, Andersen \& Lang \cite{andersen2006communities} adapted the theoretical results from \cite{spielman2004nearly} to expand a set into a community with locally minimal conductance based on lazy random walks.

However, the lazy random walk endured a much slower mixing speed and it usually took more than 500 hundred steps to converge  to a local structure compared with several steps of rapid mixing in a regular random walk.   Featuring on the seeding strategies, Whang \emph{et al.} \cite{whang2013overlapping}  established several sophisticated methods for choosing the seed set, and then used similar PageRank scheme as that in \cite{andersen2006local} to expand the seeds until a community with optimal conductance is found. Nonetheless, the performance gained by adopting these intricate seeding methods was not significantly better than that by using random seeds. This implies that a better scheme of expanding the seeds is also needed aside from a good seeding strategy. A recent work by Kloumann \& Kleinberg \cite{Isabel:KDD14} provided a systematic  understanding of variants of PageRank-based seed set expansion. They showed many insightful findings regarding the heuristics on seed set. However, the drawback of lacking a proper stop criterion has limited its functionality in practice. Even though a recently proposed heat kernel algorithm \cite{kloster2014heat} advances PageRank by introducing a sophisticated diffusion method, the detection accuracy achieved by heat kernel approach is still much lower than that of \textsc{Lemon}, which we will show in Section \ref{real_evaluation}.

 \textbf{Local spectra vs. global spectra.} Spectral methods is one of the most widely used techniques for exploratory data analysis, with applications ranging from data clustering, image segmentation to community detection etc. Spectral clustering makes use of the first few singular vectors of the Laplacian matrix associated with a graph, which are inherently global quantities and may not be sensitive to very local information. For example, in the case when provided with domain knowledge about a target region in the graph, one might be interested in finding clusters {\em only} near the specified local region in a semi-supervised manner,  which might not be otherwise well captured by a method using global eigenvectors. Therefore, in the semi-supervised setting, our pioneer work on local spectral clustering  \cite{li2015uncovering,he2015detecting}\footnote{This manuscript is an extended version of an earlier conference publication \cite{li2015uncovering}.} have substantial advantage over traditional spectral techniques, with the capability of prioritizing and learning more about a local region of the graph surrounding the seeds. Although the local spectral proposal in \cite{mahoney2012local} incorporates  the local information as an additional constraint based on the global spectral methods, the optimization program involves the entire eigenspace, which is less advantageous than using the partial invariant subspace constructed by the \emph{Krylov subspace} in our approach.

\section{preliminaries}

\subsection{Problem Statement}

 Given a network $G=(V, E)$ and a set of members $\mathcal{S}$ in the target community $\mathcal{C}$, where $|\mathcal{C}| \ll |V|$ and $|\mathcal{S}| \ll |\mathcal{C}| $,  we are interested in discovering the remaining members in $\mathcal{C}$.
Generally speaking, we focus on answering \textbf{how to accurately find a small community in time functional to the size of the community from a seed set?}

\subsection{Symbols and Definitions}
Table \ref{table:notation} summarizes a list of the different symbols we will use throughout the paper. In general, we use italic letters, e.g., $n$, $\mu$, to denote scalars; lower boldface characters, e.g. $\mathbf{y}$, to denote vectors; uppercase boldface characters, e.g., $\mathbf{A}$, to denote matrices; and script characters, e.g., $\mathcal{C}$, to denote sets. 

\begin{table}[htbp]
\begin{center}
\begin{tabular}{l|l }
\hline
\bf{Symbol}&\bf{ Definiton and description}\\
\hline
$\mathcal{S}$ & Seed set \\
$\mathcal{C}$& Detected community  \\
$\mathcal{C}^\ast$ & Ground truth community \\
$G_\mathcal{S}$ &Subgraph extracted from the neighborhood surrounding the seed set  $\mathcal{S}$\\
$N$& Size of the subgraph $G_\mathcal{S}$\\
$\mathbf{A}_\mathcal{S}$ & Adjacency matrix of subgraph $G_\mathcal{S}$ \\
$\mathbf{\bar A}_\mathcal{S}$ & Normalized adjacency matrix of subgraph $G_\mathcal{S}$ \\
$\mathbf{D}_\mathcal{S}$ & Diagonal degree matrix of subgraph $G_\mathcal{S}$ \\
$\mathbf{L}_\mathcal{S}$ & Laplacian matrix of subgraph $G_\mathcal{S}$ \\
$\mathbf{\bar L}_\mathcal{S}$ & Normalized Laplacian matrix of subgraph $G_\mathcal{S}$ \\
$\mathbf{V}_{k,l}$&$l$-dimensional local spectral subspace with $k$-step random walks.\\
$\Phi(\mathcal{V})$ & Conductance of the node set $\mathcal{V}$\\
$\lambda_i^{(\mathbf{H})}$ &The $i$-th smallest eigenvalues of matrix $\mathbf{H}$ \\
$\mathbf{y}$ & Probability indicator vector, where larger value indicates a higher\\ 
&  possibility being in the same community as the seeds\\
\hline
\end{tabular}
\end{center}
\caption{Symbols and Definitons.}
\label{table:notation}
\end{table}%

\subsection{Datasets}
{
\begin{table*}[t]
\footnotesize
\begin{center}
\begin{tabular}{l l| r r r r r r}
\hline
\bf{Domain} &\bf{Dataset} &\bf{Vertices} & \bf{Links }& \bf{Average } & \bf{Maximum} & \bf{Community}\\
 &&&&\bf{membership}&\bf{membership} & \bf{size mean}\\
 \hline
\bf{Product} &Amazon & 334,863 & 925,872 & 0.11 & 49 & 39\\
\bf{Collaboration} & DBLP &317,080 &1,049,866  & 0.22 & 11 & 251\\
\bf{Social} &YouTube & 1,134,890 & 2,987,624 & 0.05 & 41 & 79 \\
\bf{Social}& Orkut & 3,072,441  & 117,185,083&9.56 & 504 & 83\\
\hline
\end{tabular}
\end{center}
\caption{Statistics for the real networks.}
\label{real_data_summary}
\end{table*}%
}
\subsubsection{Synthetic datasets}

The LFR benchmark graphs \cite{lancichinetti2008benchmark} have been widely adopted for the purpose of evaluating the performance  of community detection algorithms.
LFR datasets are generated with built-in community structure that resembles the features found in most real-world networks with power-law degree distribution. It provides researchers with rich flexibility to control the network topology by tuning different parameters, including the graph size $n$, the average degree $\bar k$, the maximum degree $k_{max}$, the minimum and maximum community size $|\mathcal{C}|_{min}$ and $|\mathcal{C}|_{max}$, the mixing parameter $\mu$, the overlapping membership $om$ and the number of vertices with overlapping membership $on$. Among these parameters, the mixing parameter $\mu$ has the most significant impact on the network topology, which controls the fraction of links for each vertex that cross to a community with which the vertex is not associated. Usually, larger $\mu$ would result in lower detection accuracy.

Xie et al. \cite{xie2013overlapping} have performed a thorough performance comparison of different state-of-the-art overlapping community detection algorithms on LFR benchmark datasets. To make the performance evaluation of our algorithm consistent  with that in \cite{xie2013overlapping}, we adopt the same parameters in our paper. In total, we generate two sets of networks with mixing parameter $\mu=0.1$ and $\mu=0.3$  respectively.

We vary the parameter $om$ from $2$ to $8$ for each $\mu$ and obtain a total of $14$ networks.
Table \ref{LFR_summary} lists the value of the parameters we have used for generating the LFR datasets.

\begin{table}[htbp]
\begin{center}
\small
\begin{tabular}{l|l |l}
\hline
\bf{Parameter}&\bf{ Description}& \bf{Value}\\
\hline
$n$& graph size & 5000\\
$\mu$ & mixing parameter & \{0.1, 0.3\}\\
$\bar k$ &average degree & 10\\
$k_{max}$ & maximum degree & 50\\
 $|\mathcal{C}|_{min}$ &minimum community size &$20$\\
 $|\mathcal{C}|_{max}$&maximum community size &$100$\\
$\tau_1$&node degree distribution exp.& $2$\\
$\tau_2$ & community size distribution exp.& $1$\\
$om$ &overlapping membership & \{2, 3, ..., 8\}\\
 $on$& overlapping node &2500\\
\hline
\end{tabular}
\end{center}
\caption{Parameters for the LFR datasets.}
\label{LFR_summary}
\end{table}%

\subsubsection{Real datasets}

For the purpose of testing on real networks, we include four datasets with ground truth community membership from Stanford Network Analysis Project\footnote{\url{http://snap.stanford.edu}}. These datasets span various domains of network applications, including product networks (Amazon), collaboration networks (DBLP), and online social networks (YouTube and Orkut)\footnote{For all the four real datasets, we adopt the top $5000$ communities that possess the highest quality according to \cite{Jaewon:ICDM2012}.}. Each of the networks can be viewed as an undirected, connected graph.

 The statistical information of the datasets is summarized in Table \ref{real_data_summary}.

\subsection{Evaluation Metric}

For the evaluation metric, we adopt F1 score to quantify the similarity between the algorithmic community $\mathcal{C}$ and the ground truth community $\mathcal{C}^*$. The F1 score for each pair of $(\mathcal{C},\mathcal{C}^*)$ is defined by:
\begin{equation}
F_1(\mathcal{C},\mathcal{C}^*) = \frac{2 \cdot Precision(\mathcal{C},\mathcal{C}^*) \cdot Recall(\mathcal{C},\mathcal{C}^*) }{Precision(\mathcal{C},\mathcal{C}^*) +Recall(\mathcal{C},\mathcal{C}^*) },
\end{equation}
where the precision and recall are defined as:
\begin{equation}
Precision(\mathcal{C},\mathcal{C}^*) = \frac{|\mathcal{C}\cap\mathcal{C}^*|}{|\mathcal{C}|},
\end{equation}
\begin{equation}
Recall(\mathcal{C},\mathcal{C}^*) = \frac{|\mathcal{C}\cap\mathcal{C}^*|}{|\mathcal{C}^*|}.
\end{equation}

Throughout the paper, unless otherwise pointed out, the experimental results on synthetic data for each instance are given by the statistical mean and standard deviation based on 24 test cases\footnote{Each local expansion process from a seed set can be viewed as a test case.}; and the experimental results on real datasets for each instance are based on 120 test cases.  All the ground truth communities for testing are randomly chosen.
The randomness of batch tests can guarantee the elimination of statistical bias in our tests.

\section{Local expansion via minimizing one norm}

\subsection{Algorithm Overview}

Spectral clustering makes use of a small number of singular vectors proportional to the number of communities in the network. If a graph has thousands of small communities, it is impractical to calculate a number of singular vectors greater than the number of communities. We are experimenting with a fundamentally new technique, which 
does not require the burdensome computation of a large number of singular vectors. Before explaining our local spectral approach for finding overlapping communities, it is necessary to make  clear what we mean by \emph{local spectra}.

In traditional spectral clustering methods, one finds the first few singular vectors of the Laplacian matrix\footnote{In the literature, several different definitions of graph Laplacian exist. Readers can refer to \cite{ng2002spectral} for more details, which serves as a good introductory paper on spectral clustering. } of a graph $G$ with $n$ vertices.   Suppose the first $d$ singular vectors are obtained, one can form an $n \times d$ matrix as a latent space. Then one associates with each vertex a point in this latent space whose coordinates are given by the entries of the corresponding row in the matrix. Vertices are clustered using some method such as $k$-means clustering algorithm.  This method is not likely to work well if the communities are small and heavily overlapping with each other.

We make two fundamental changes to this method. The first modification is to overcome the drawback of computing the singular vectors. Intuitively, the vertices around the seed members are more likely to be in the target community,  thus a random walk serves as a natural subroutine to reveal these potential members. 

We start a random walk from several known members in the target community and run for a few steps. The number of random walk steps should be long enough to reach out to the vertices in the target community, but not long enough to spread out to the entire graph. Instead of  considering a single probability vector, we consider the span of a few dimensions of vectors after the short random walks and use it as the approximate invariant subspace (\emph{local spectra}).  The second is to handle the overlapping situation. Instead of using $k$-means to partition the points in the latent space into disjoint clusters, we look for the minimum $0$-norm vector  in the span of the invariant subspace obtained above, such that the seed members are in its support. We want to find rows in the invariant subspace that point in nearly the same direction as seed members. We will use $1$-norm vector as a proxy for the minimum $0$-norm vector since finding the $0$-norm vector is an NP-hard problem.

In the following, we give a formal description of our local spectral approach {\textsc{Lemon}} for detecting target communities from a small seed set.
Given the input of a set of few vertices $\mathcal{S}$ that are already known to be in the target ground truth community $\mathcal{C}^*$, our algorithm would output the algorithmic community $\mathcal{C}$ such that the F1 measure for scoring the similarity between $\mathcal{C}$ and $\mathcal{C}^*$ is maximized. Note that the each seed set expansion is operated on a small sampled graph $G_\mathcal{S} = (\mathcal{V}_\mathcal{S}, \mathcal{E}_\mathcal{S})$, extracted from the neighborhood surrounding the seed set $\mathcal{S}$. The details of sampling local graph will be given in Section \ref{sample}.\\

{\bf Step 1. Generate the local spectra:}

 Consider the subgraph graph $G_\mathcal{S}$ extracted from the neighborhood surrounding the seed set $\mathcal{S}$.  Let $ \mathbf{ \bar A_\mathcal{S}}=\mathbf{D_\mathcal{S}}^{-1/2}(\mathbf{A_\mathcal{S}}+\mathbf{I}) \mathbf{D_\mathcal{S}}^{-1/2}$ be the normalized adjacency matrix of the graph.
 We define  the normalized adjacency matrix $\mathbf{ \bar A_\mathcal{S}}$ of the graph $G_\mathcal{S}$ as 
\begin{equation}\label{eqn:normalized_matrix}
\mathbf{ \bar A_\mathcal{S}} \myeq \mathbf{D_\mathcal{S}}^{-1/2}(\mathbf{A_\mathcal{S}}+\mathbf{I}) \mathbf{D_\mathcal{S}}^{-1/2},
\end{equation}
where $\mathbf{A_\mathcal{S}}$ and $\mathbf{D_\mathcal{S}}$ denotes the adjacency matrix and the diagonal degree matrix of $G$, respectively. Consider a random walk starting from exemplary vertices in $\mathcal{S}$. Let $\mathbf{p_0}$ denote the initial probability vector where the total probability is evenly distributed among the seed members. We describe how to efficiently construct the local spectra by iteratively transforming the orthonormal basis starting with a {\em Krylov subspace} defined below. 

\begin{mydef}
The order-$l+1$ Krylov subspace generated by the matrix $\mathbf{A}$ and vector $\mathbf{p}_0$ is the linear spanned subspace defined by the probability vectors in $l$ successive random walks
\end{mydef}
\begin{equation}
\mathcal{K}_{l+1}(\mathbf{A},\mathbf{p}_0) = {\text{span}}~\Big(\mathbf{p}_0,\mathbf{A} \mathbf{p}_0,...,\mathbf{A}^l \mathbf{p}_0 \Big)
\end{equation}

\begin{figure}[t]
\centering
\includegraphics[width=\columnwidth]{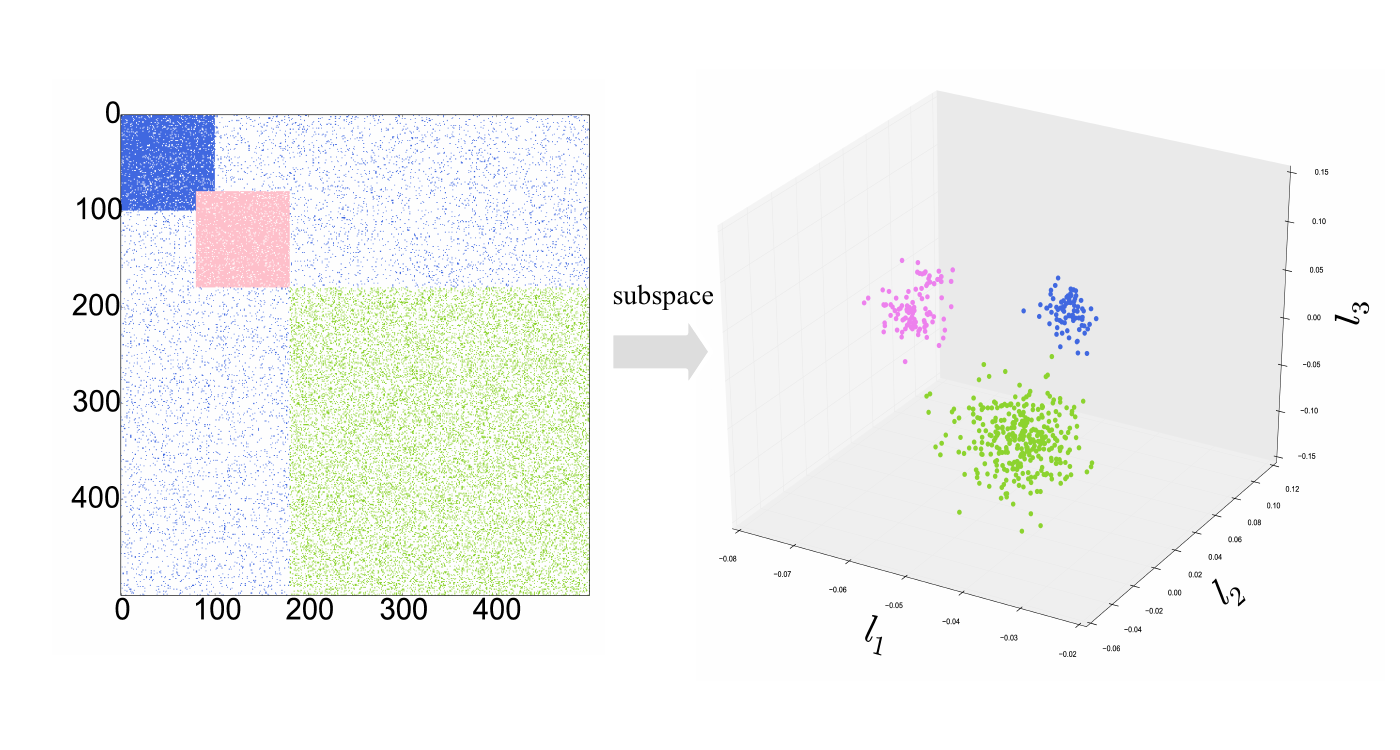}%
\caption{{ An example of local spectral subspace $\mathbf{V}_{3,3}$. The synthetic subgraph $G_s$ is generated with Erd\H{o}s-R\'{e}nyi $G(n,p)$ model with background noise $p=0.05$. The spammer group $A$ and $B$ (denoted by blue and pink respectively) are of size 100 with edge probabolity $p=0.9$, with partial overlapped 20 nodes. The non-spammer group $C$ (denoted by the green color) has size 320 with $p=0.2$. The subspace is generated by Algorithm 1 starting from the seed with index 10 in the spammer group $A$.}}%
\label{subfiga}%
\end{figure}

In Algorithm 1, we briefly summarize the procedure of calculating the local spectral subspace {\footnote{In the experiments on real datasets, we fix the walk step $k$ and dimension $l$ to be 3 and 3 respectively. For LFR benchmark datasets, we adopt all together 6 combinations for the (step, dimension) tuple: $(2,3),(2,4),(2,5),(3,3),(3,4),(3,5)$ and the highest F1 score among these combinations will be returned. }} from a specified seed set $\mathcal{S}$. We start by calculating the initial invariant subspace $\mathbf{V}_{0,l}$, which is the orthonormal basis of $\mathcal{K}_{l+1}(\mathbf{A_\mathcal{S}},\mathbf{p}_0)$. And the local spectral subspace can be then obtained by iterating the process specified in \textsc{Line} 4-6 of Algorithm 1. Figure \ref{subfiga} shows an example local spectral subspace $\mathbf{V}_{3,3}$, generated from a synthetic graph with  Erd\H{o}s-R\'{e}nyi $G(n,p)$ model. 

\floatname{algorithm}{Algorithm}
\renewcommand{\algorithmicrequire}{\textbf{Input:}}
\renewcommand{\algorithmicensure}{\textbf{Output:}}

\begin{algorithm}[htbp]
  \caption{\textsc{Local}\textsc{Spectral}($G_\mathcal{S},\mathcal{S}$)}
  \begin{algorithmic}[1]
    \Require{subgraph $G_\mathcal{S}$, subspace dimension $l$, and random walk step $k$}
    \Ensure {local spectra $\mathbf{V}_{k,l}$}
    \State Compute normalized adjacency matrix $\mathbf{ \bar A_\mathcal{S}}$ using (\ref{eqn:normalized_matrix})
    \State Initialize $\mathbf{p}_0$
    \State $\mathbf{V}_{0,l} = \mathtt{orth} (\mathcal{K}_{l+1}(\mathbf{ \bar A_\mathcal{S}},\mathbf{p}_0))$
    \For {$i = 1,...,k$}
    \State $\mathbf{V}_{i,l}\mathbf{R}_{i,l}=\mathbf{\bar A_\mathcal{S}} \mathbf{V}_{i-1,l}$
    \Comment {$\mathbf{R}_{i,l}\in \mathbb{R}^{n\times l}$ is obtained by QR factorization so that $\mathbf{V}_{i,l}$ is orthonormal.} 
    \EndFor
\State {\bf Return} {local spectra $\mathbf{V}_{k,l}$}
  \end{algorithmic}
\end{algorithm}

{\bf Step 2. Seek for a sparse vector}

With the local spectra $\mathbf{V}_{k,l}$, we solve the following linear programming problem,
\begin{align*}
\min &~~~ ||\mathbf{y}||_1 \\
\text{s.t.} & ~~~\mathbf{y}\in \text{span} (\mathbf{V}_{k,l}), \\
&~~~\mathbf{y} \ge \mathbf{0},\\
&~~~ \mathbf{y}(\mathcal{S}) \ge 1,
\end{align*}

where the first constraint indicates that $\mathbf{y}$ is in the space of $\mathbf{V}_{k,l}$.
The element in $\mathbf{y}$
 indicate the likelihood for the corresponding vertex belong to the target community, which is non-negative.
The third constraint enforces that seeds are in the support of sparse vector $\mathbf{y}$.

After sorting the elements in $\mathbf{y}$ in non-ascending order and  getting a vector $\mathbf{\hat y}$,
 the vertices corresponding to the top $|\mathcal{C}|$ elements in $\mathbf{\hat y}$ are returned as the detected community with respect to the seed set $\mathcal{S}$. \\

{\bf Step 3. Reseeding}

Augment the initial seed set by adding the vertices corresponding to the top $t$ elements of $\mathbf{\hat y}$. Denote the augmented seed set as $\mathcal{S}'$. Then repeat step 1 and step 2 using the augmented seed set $\mathcal{S}'$. The detection accuracy can be improved through iterations via increasing $t$ by a constant number $s$ each time. We define $s$ to be the seed expansion step, which is used as a tunable parameter for adjusting the convergence rate. Usually, the larger expansion step would result in lower performance but a faster running speed with less iterations. In the experiments, we fix the seed expansion step to be 6 for both synthetic and real datasets.
The number of iterations for the seed expansion is determined by the stop criteria (Section \ref{stop}).\\

\subsection{Parameter Sensitivity}\label{parameter}

The random walk step $k$ and subspace dimension $l$ are the key parameters in the local spectral clustering algorithm.
We conduct  parameter sensitivity study for these two parameter on the four real datasets.


\subsubsection{Subspace dimension}

To study the parameter of subspace dimension $l$, we fix the random walk step to be 3, and vary the number of dimension $l$ from 1 to 15.
Figure \ref{LFR_para} (left panel) shows that  changing the dimension $l$ does not cause significant fluctuation of the performance. On one hand, choosing a large dimension $l$  is undesirable because it would increase the computation cost in the step of generating local spectra. On the other hand, when dimension degrades to $l=1$, the standard deviation of F1 score becomes significant, making the detection accuracy unstable. In this paper, we fix $l=3$ because the experiment suggests that setting $l=3$ can statistically achieve both high and stable performance. Note that such observation holds not only for Amazon network, but  for the remaining real datasets as well.


\subsubsection{Random walk step}

To investigate how the step of random walk affects the algorithm performance, we fix the dimension $l$ to be 3, and vary the random walk step $k$ from 1 to 15.
Figure \ref{LFR_para} (right panel) shows that the average F1 score plateaus as $k$ increases, and 3-step random walk  can yield the algorithm's full potential. The standard deviation, however, significantly increases when $k$ exceeds 10.  This indicates that longer random walk is undesirable for stably uncovering the local community structure. Throughout the paper, we fix the random walk step $k=3$ for the real datasets\footnote{For LFR benchmark graphs, we adopt all together 6 combinations for the (step, dimension) tuple: $(2,3),(2,4),(2,5),(3,3),(3,4),(3,5)$ and return the highest F1 score among using these combinations.
}.

\begin{figure*}[htbp]
\begin{center}$
\begin{array}{cc}
\includegraphics[width=0.45\linewidth]{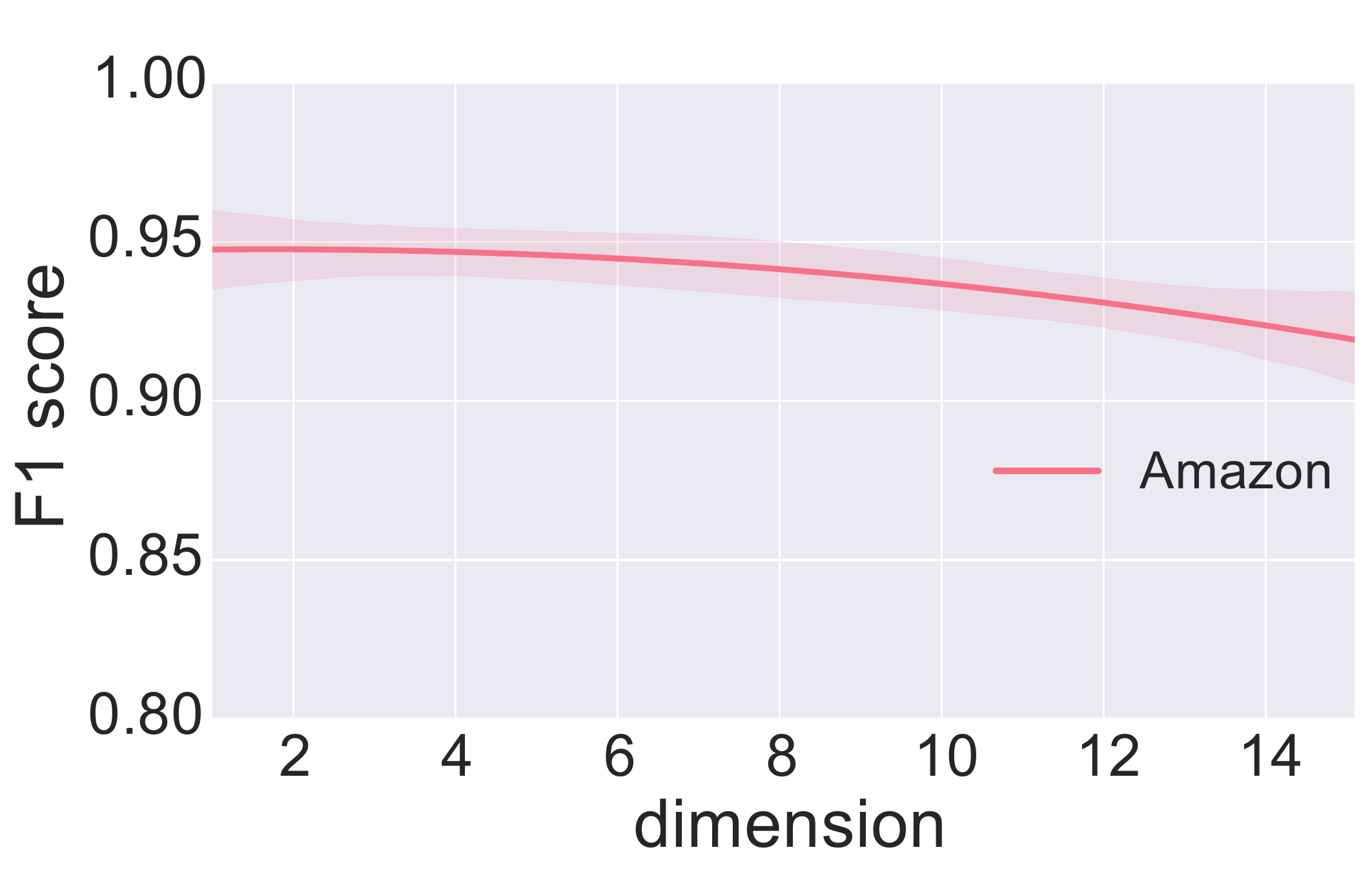}     \qquad\qquad\qquad      &  
\includegraphics[width=0.4\linewidth]{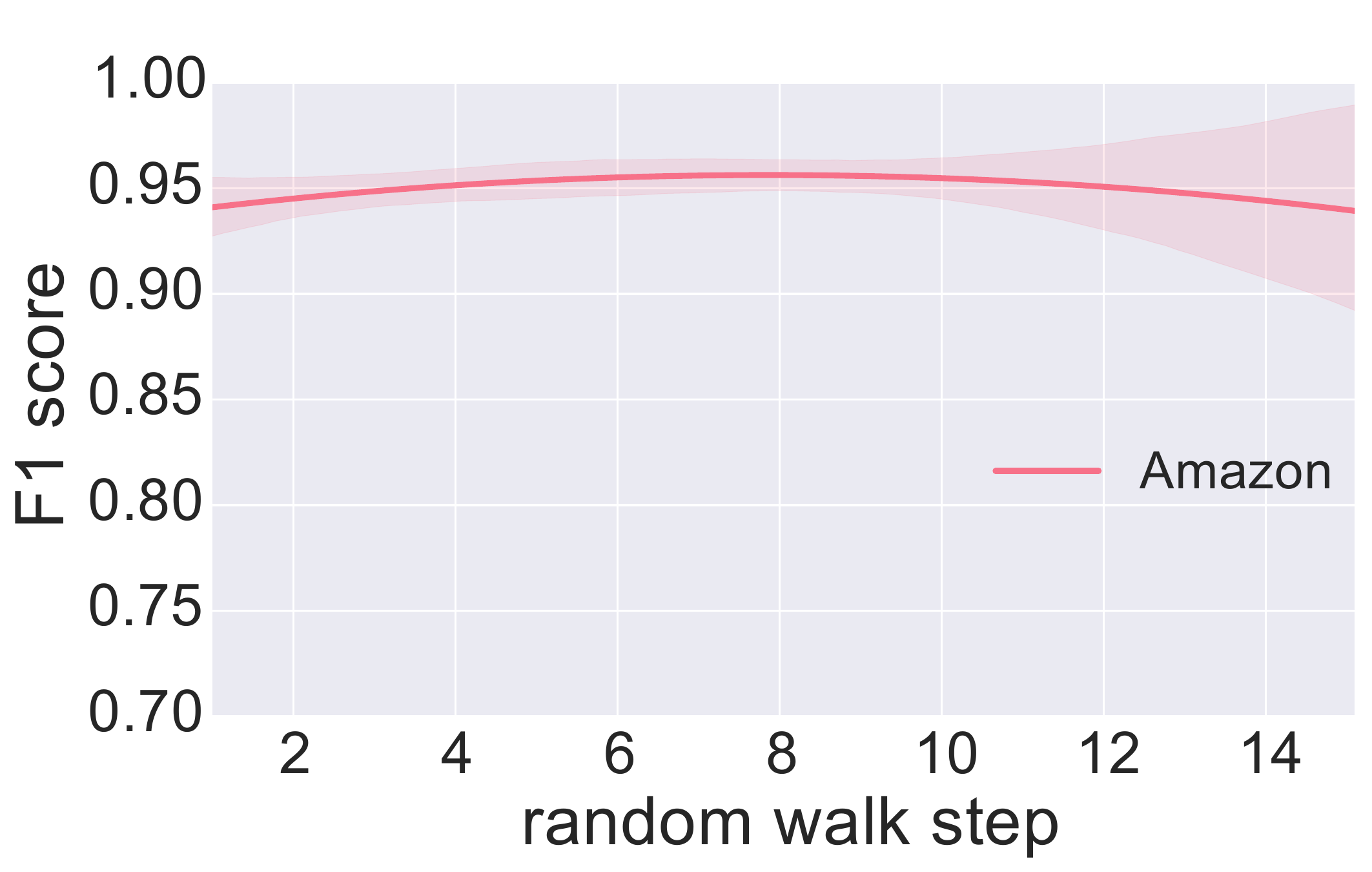} \qquad\qquad\qquad
\end{array}$
\end{center}
\caption{{{The average F1 score on Amazon network with varying dimensions $l$ and random walk step $k$, respectively. The plots depict the statistical regression line with a 95\% confidence interval.}}}
\label{LFR_para}
\end{figure*}

\begin{figure*}[htbp]
\begin{center}$
\begin{array}{cc}
\includegraphics[width=0.45\linewidth]{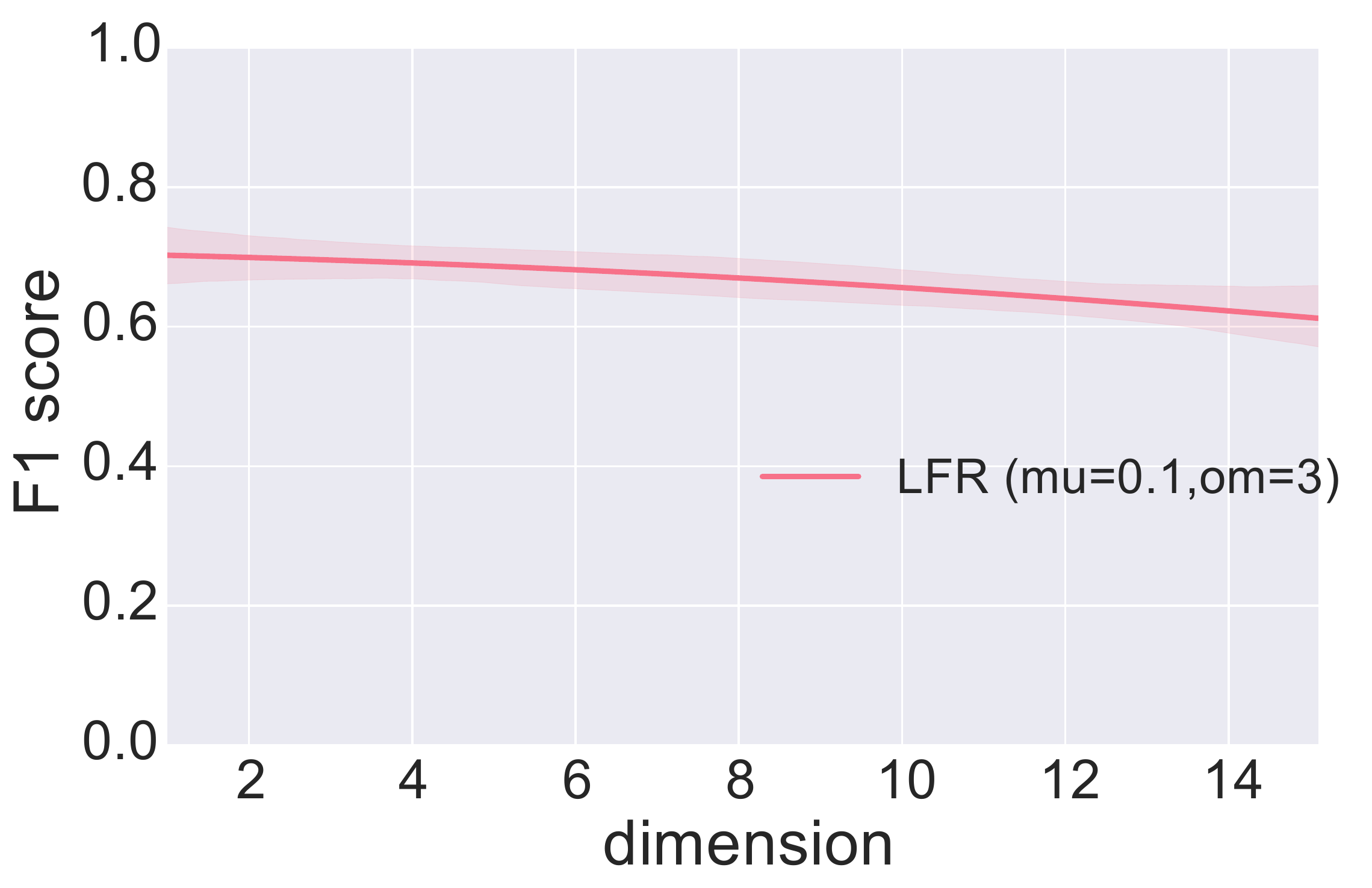}     \qquad\qquad\qquad      &  
\includegraphics[width=0.4\linewidth]{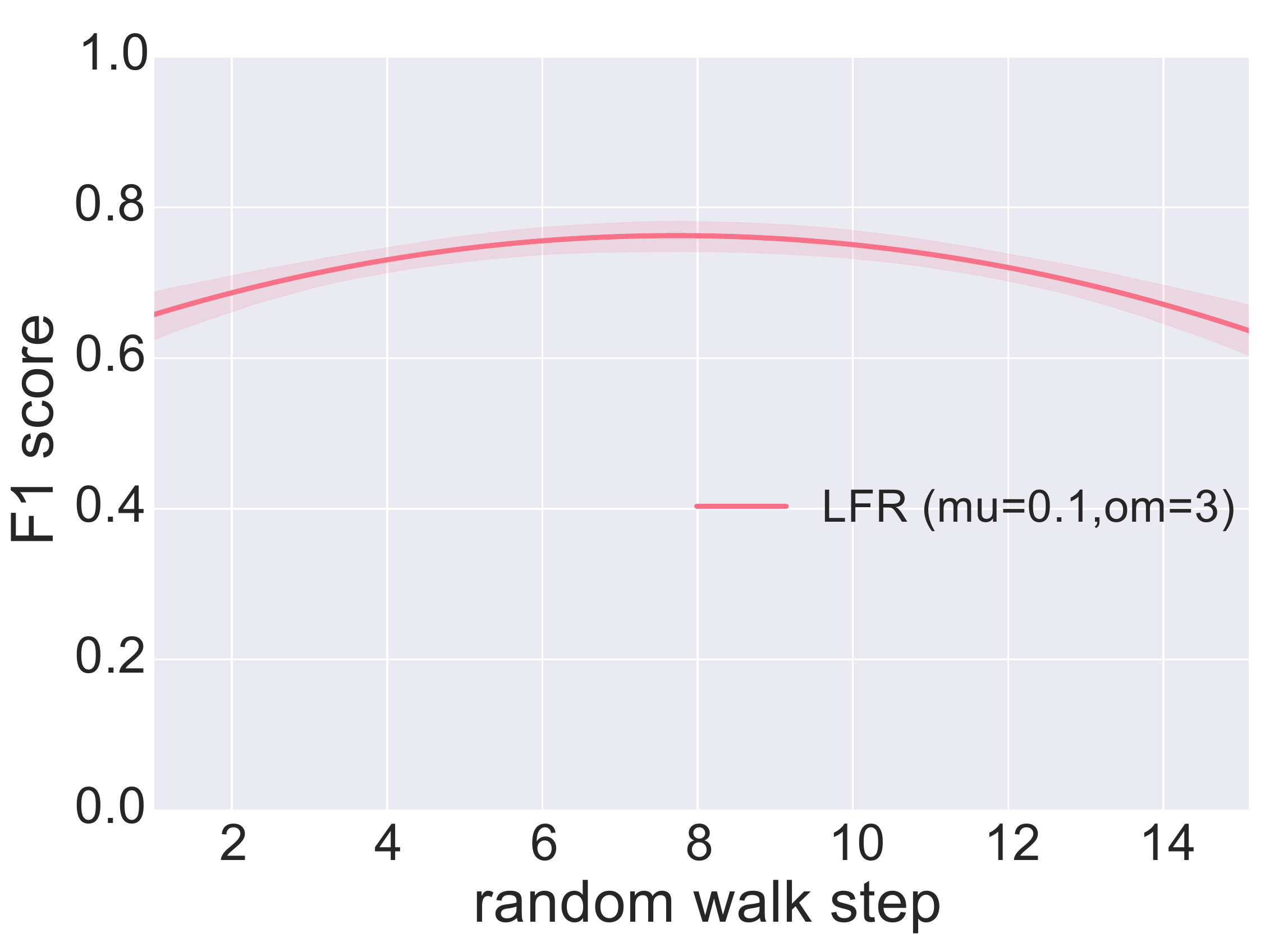} \qquad\qquad\qquad
\end{array}$
\end{center}
\caption{{{The average F1 score on LFR benchmark graph ($\mu=0.3$) with varying dimensions $l$ and random walk step $k$, respectively. The plots depict the statistical regression line with a 95\% confidence interval.}}}
\label{LFR_para}
\end{figure*}

\subsection{Local Spectra vs. PageRank} \label{compare_pr}

The local spectra clustering approach and PageRank algorithm both utilize short random walks to detect the local community structure. PageRank is solely based on the single probability vector, and the latent community members are selected through ranking the probability value among vertices. The local spectral clustering advances PageRank-like algorithms by forming a subspace based on the short random walk, and seeking for a sparse vector such that the seeds are in its support.

\begin{table}[htbp]
\begin{center}
\begin{tabular}{l | l   l  l l}
\hline
 & \bf{Amazon} & \bf{DBLP} & \bf{YouTube} & \bf{Orkut}\\
\hline
\bf{LEMON} & 0.953  & 0.665 &0.240 &0.202   \\
\bf{PageRank}& 0.140 & 0.115 & 0.136 & 0.044\\
\hline
\end{tabular}
\end{center}
\caption{{Comparison of the mean F1 score with local spectral clustering and PageRank. LEMON' is to use the ground truth size to decide the community size,
and LEMON is to determine the community size automatically by the first local minimal of the conductance value.)}}
\label{pagerank}
\end{table}%

By comparing the performance of these two approaches on the real datasets,  we show that seeking for the sparse vector is more effective than directly sorting the probability vector alone. Table \ref{pagerank} shows the comparison of average F1 score obtained by local spectral clustering and PageRank, respectively.\footnote{The statistical results of PageRank algorithm is sourced from \cite{kloster2014heat}.} From the result, we see that the performance gain brought by the local spectral method is significant, where it achieves more than 5 times higher accuracy on Amazon, DBLP and Orkut networks. We also take into account of a variety of state-of-the-art community detection algorithms for performance comparison  in Section \ref{compare}.

\begin{figure}[htbp]
\begin{center}$
\begin{array}{cc}
\includegraphics[width=0.6\linewidth]{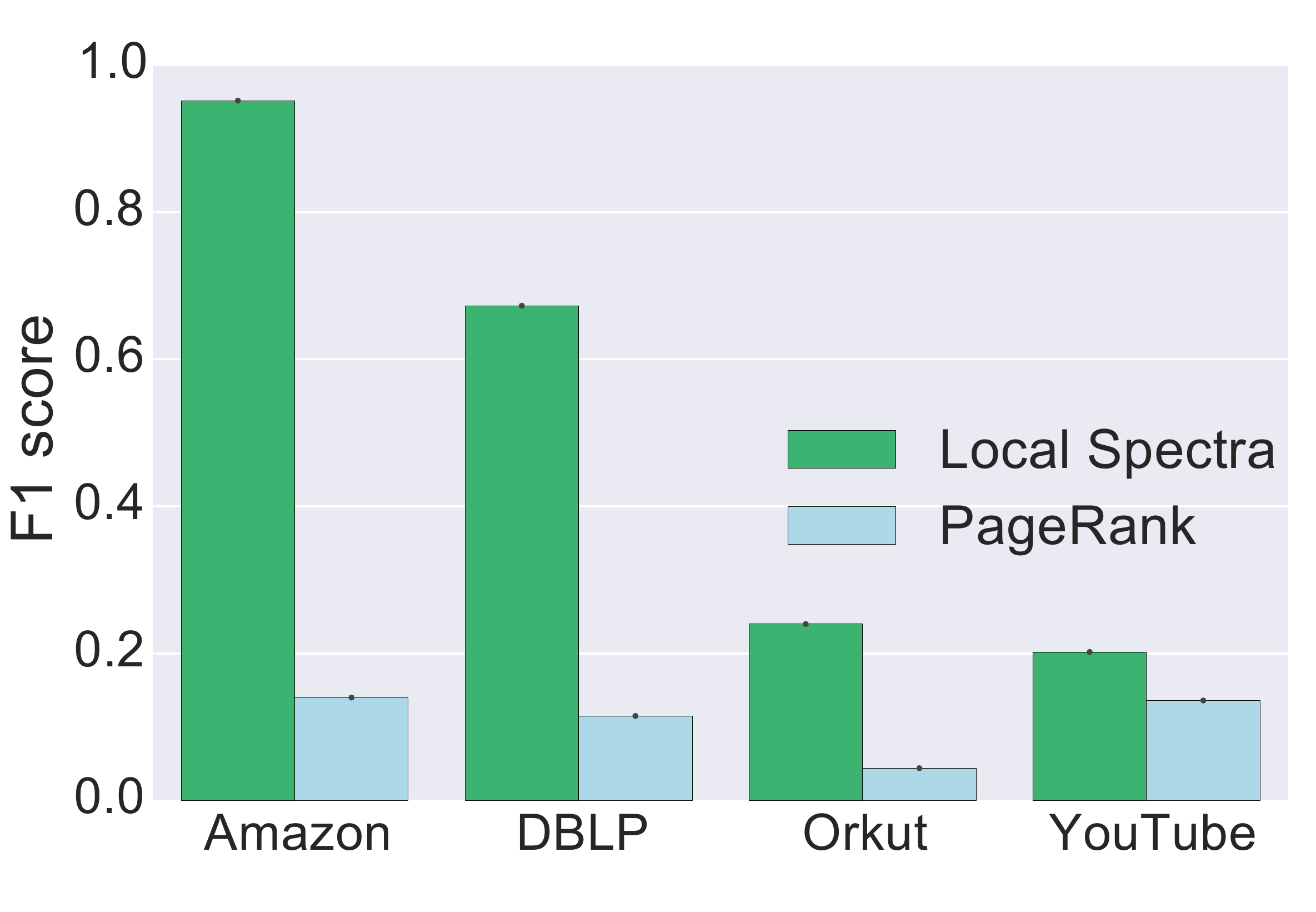}
\end{array}$
\end{center}
\caption{{Comparison of the average F1 score with local spectral clustering and PageRank algorithms. Height of the bars shows the mean and 95\% confidence interval.}}
\label{LFR_naive}
\end{figure}

\subsection{Complexity Reduction by Sampling Method}\label{sample}


 If one wants to uncover a small community within a large network consisting of billions of vertices, it would be very costly to take all the vertices into account. We want to discover the target community accurately  while keeping the number of vertices examined small. Sampling method can effectively solve the memory consumption issue when one wants to find a local community  within a large graph, since the whole graph does not have to be stored in memory.

In practice, the unknown members in the target community are more likely to be around the seed members, and are usually a few steps away from the seeds. This observation motivates us to reduce the complexity by taking only a portion of the graph into consideration. Ideally, this partial graph should contain as many vertices in the target community as possible, and  maintains a small size of the same scale as that of the target community.

To sample the graph, we expand the seed set using random walk. After a few steps of the random walk, vertices with large probability are more likely to be in the target community while vertices with small probability being reached would be treated as redundant ones. If the target community exists for the seed set, then according to \cite{andersen2006communities}, this target community would serve as a bottleneck for the probability to be spread out. 
It is worthwhile noting that other expansion methods such as breadth-first-search (BFS) would entirely ignore the bottleneck defining the community and rapidly mix with the entire graph before a significant fraction of vertices in the community have been reached. The subgraph returned by BFS usually contains less vertices in the target community than the subgraph of the same size obtained by random walk technique.

\begin{figure*}[t]
\begin{center}$
\begin{array}{cc}
\includegraphics[width=0.45\linewidth]{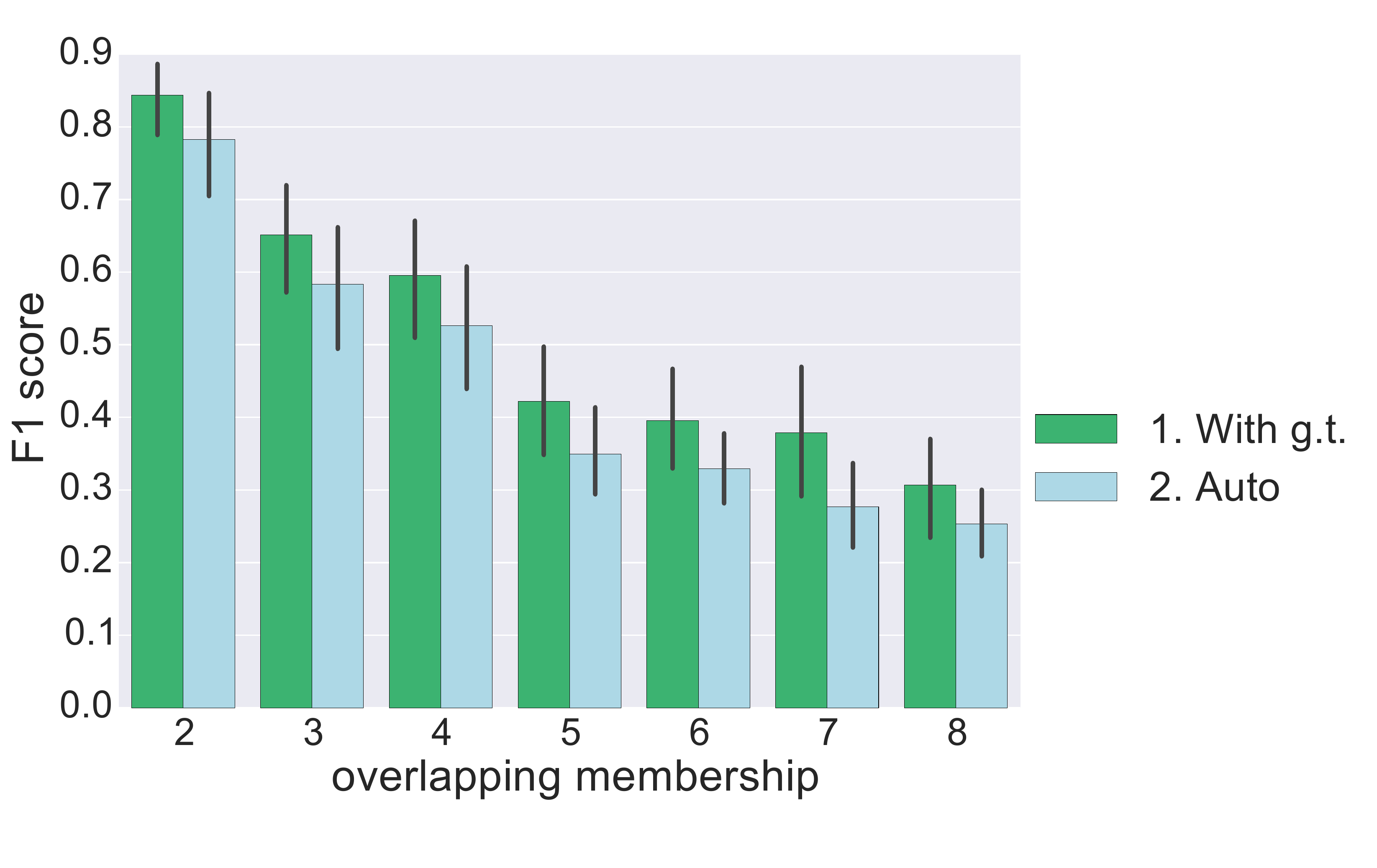} &
\includegraphics[width=0.45\linewidth]{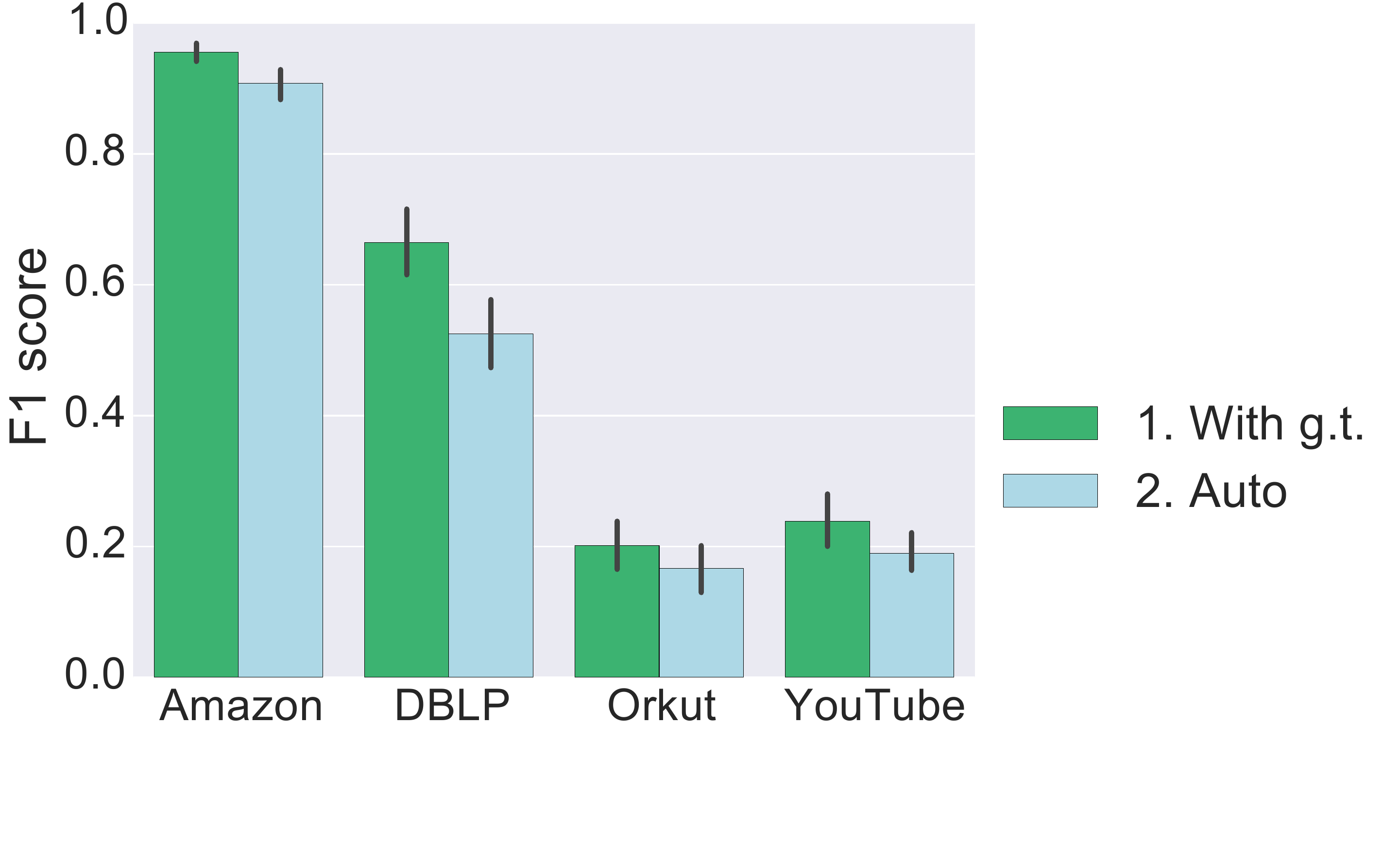}
\end{array}$
\end{center}
\caption{{Comparison of the average F1 score with ground truth and automatic size determination. The left corresponds to the LFR datasets when $\mu=0.3$ and the right  corresponds to the real datasets. }}
\label{LFR_auto}
\end{figure*}

In the experiments on real datasets, we conduct a random walk starting from the seed set until the probability has been spread out to $\alpha\cdot|\mathcal{C}|_{\text{avg}}$ vertices, where $\alpha$ is some constant and $|\mathcal{C}|_{\text{avg}}$ is the average community size in the graph\footnote{A fast implementation method for updating the probability vector of the random walks is featured in detail in \cite{andersen2006communities}, Section 4.}. 
Note that $\alpha\cdot|\mathcal{C}|_{\text{avg}}$ should be large enough to be able to cover as many vertices in the ground truth community as possible.

This newly obtained subgraph will be used for the remaining computation. The complexity of our algorithm now depends on the size of the  subgraph after sampling, which is $O({|\mathcal{C}|_{\text{avg}}}^\tau)$ for some small constant $\tau$.

\begin{table}[htbp]
\begin{center}
\begin{tabular}{l | r  r  r  r r}
\hline
\bf{Dataset} & \bf{Coverage} & \bf{Sample }& \bf{$|\mathcal{C}|_{\text{avg}}$} & \bf{Subgraph}\\
& \bf{ratio}& \bf{rate} &  &  \bf{size} \\
\hline
\bf{Amazon}    & 1.00 & 0.0087  &  39   & 2913  \\
\bf{DBLP}      & 0.98   & 0.0076  &  251      &  2409 \\
\bf{YouTube}   & 0.66 & 0.0033  &  79   & 3745  \\
\bf{Orkut}     & 0.64   & 0.0011  &  83      & 3379 \\
\hline
\end{tabular}
\end{center}
\caption{{Statistics of the mean values for the sampling method on real datasets.}}
\label{real_sample}
\end{table}%

Table \ref{real_sample} gives the statistics after applying sampling method to the real networks.

For example, in DBLP network, setting $\alpha$ to be around 10 would yield a subgraph containing on average $98\%$ vertices in the ground truth community. After sampling, we only need to deal with a subgraph of size around $2400$ instead of 317,080, bringing a significant reduction of both  temporal and spatial complexity.

\subsection{Round Diffusion Vector via Sweeping Cut}\label{stop}

If there are ground truth communities available, the above algorithm is guaranteed to stop within few iterations since the seed set will no longer augment once  its size  exceeds that of the ground truth community. The algorithm would then return the community found with the highest F1 score during the iterations as the result.
However, in real case, we don't know the exact size of the communities,
causing the ambiguity for most locally based detection algorithm to decide when is the proper time to terminate expanding such that the discovered community is a ``good" community. It is thus important to solve the two issues: 1) how to automatically determine the size of the community given a seed set $\mathcal{S}$, and 2) when to stop growing the seed set during the reseeding process.

\subsubsection{Determine the size of the community}

 It has already been shown that random walks produce communities with conductance guarantees and ensure a small boundary defining a natural community in locally based detection algorithms \cite{andersen2006communities}.  The intuition is that adding irrelevant vertices to the target community would inevitably cause the conductance to increase,
  and finding a low-conductance community could ensure the closeness between the detected members and the known seed set. 
  
    A commonly adopted method of rounding the diffusion values into labels is to perform a sweep-cut procedure on the nodes ranked by the diffusion value, with an objective of minimizing the graph cut metric such as {\em conductance} \cite{andersen2006local,mahoney2012local,whang2013overlapping}. As we will see, the local conductance for a small group of vertices in the graph contains valuable information and enables us to design effective stopping criteria for our algorithm. We define conductance using the generalized  Rayleigh quotient specified below:

\begin{mydef}
Let $\mathbf{x}\in \{0,1\}^N$ denote the binary indicator vector for the subset $\mathcal{\tilde V}\subseteq \mathcal{V}_s$ and $\mathbf{H}\in \mathbb{R}^{N\times N}$ is any symmetric matrix. The Rayleigh quotient with respect to $\mathbf{H}$ is expressed as the quadratic form of 
\begin{equation}\label{eq:rayleigh}
R_{\mathbf{H}}(\mathbf{x}) =\frac{\mathbf{x}^T\mathbf{H}\mathbf{x}}{\mathbf{x}^T\mathbf{x}}
\end{equation}
\end{mydef}
In particular, conductance of the set $\mathcal{\tilde V}$ measures the fraction of edges leaving $\mathcal{\tilde V}$ among all the edges incident on $\mathcal{\tilde V}$, and can be expressed using a generalized Rayleigh quotient 
\begin{equation}
\Phi(\mathcal{\tilde V}) = R_{\mathbf{L_\mathcal{S}},\mathbf{D_\mathcal{S}} }(\mathbf{x}) =\frac{\mathbf{x}^T\mathbf{L_\mathcal{S}}\mathbf{x}}{\mathbf{x}^T\mathbf{D_\mathcal{S}}\mathbf{x}} = \frac{\mathbf{x}^T(\mathbf{D_\mathcal{S}}-\mathbf{A_\mathcal{S}})\mathbf{x}}{\mathbf{x}^T\mathbf{D_\mathcal{S}}\mathbf{x}}, 
\end{equation}
where $\mathbf{L_\mathcal{S} }= \mathbf{D_\mathcal{S}}-\mathbf{A_\mathcal{S}} $ is the Laplacian matrix of graph $G_{\mathcal{S}}$.

Now suppose we have a rough estimation of the lower and upper bound for the size of communities in a graph, which we denote by $|\mathcal{C}|_{\min}$ and $|\mathcal{C}|_{\max}$ respectively. We could modify the original algorithm in the following way.

 At step 2, after obtaining the sorted sparse vector $\mathbf{\hat y}$, we are hoping to truncate the sorted vector at some point $y_g$ such that all the vertices corresponding to the elements no less than $y_g$ are included in the algorithmic community.  The crux lies in that we do not know which is the best position to truncate the vector $\mathbf{\hat y}$. To solve this issue, we denote $\Lambda_i$ as the set of vertices corresponding to the top $i$ elements in $\mathbf{\hat y}$. We then sweep over the sets from  $\Lambda_{|\mathcal{C}|_{\min}}$ to  $\Lambda_{|\mathcal{C}|_{\max}}$ and calculate the corresponding conductance for each of the sets.  In practice, the value of the conductance with respect to varying size would usually change in a non-monotonic pattern that decreases first and then  increases later on. We then adopt the minimum conductance encountered on this curve as the estimated size of the community with respect to the seed set $\mathcal{S}$, which we denote by $\Phi_{\mathcal{S}}^{\min}$.

\subsubsection{Stop the reseeding process}

As we keep augmenting the seed set through reseeding at step 3, a different seed set would result in a different sparse vector $\mathbf{\hat y}$ and thus lead to potentially different algorithmic communities. Practically, one of these seed sets during the augmenting process would achieve the highest F1 score.  And it remains to address the issue of when to stop growing the seed set so that it finds the community that resembles most of the ground truth community. This issue can be solved in a similar fashion as that for determining community size. Specifically, we keep track of the value of  $\Phi_{\mathcal{S}}^{\min}$ for different seed set during the expansion, and stop to grow the seed set when $\Phi_{\mathcal{S}}^{\min}$ reaches a local minimum and starts to increase for the first time.

\subsubsection{Auto detect size vs ground truth size}

To verify our method, we compare the performance after applying the stop criteria with that obtained using ground truth communities. Figure \ref{LFR_auto}

shows the statistical result of F1 score on both synthetic and real datasets. On both datasets, the F1 score with automatic size determination is only lowered by 10\% on average compared with the performance with available ground truth. This implies that our method is applicable for finding communities that mostly resemble the ground truth communities on both synthetic and real datasets in different domains.  It also suggests that our method can be applied in practice to uncover natural communities  in the situation when no ground truth is available.

\subsection{Bounding the Performance}

In the following discussion, we bound the measure of ``tightness" of the extracted community with respect to the subgraph $G_\mathcal{S}$ by relating spectral properties to Rayleigh quotients. We start by providing several theorems and lemmas that will be used for deriving the bound of conductance. 

\begin{mytheo}\label{theo:cheeger}
(Cheeger's Inequality) Let $\lambda_2$ be the second smallest eigenvalue of the Laplacian matrix for a graph $G_{\mathcal{S}}$. Then $\phi(G_{\mathcal{S}}) \ge \frac{\lambda_2}{2}$, where $\phi(G_{\mathcal{S}}) = \min_{\mathcal{\tilde V} \subseteq \mathcal{V_{\mathcal{S}}}} \Phi(\mathcal{\tilde V})$.
\end{mytheo}
There are many proofs known for this theorem \cite{chung1997spectral}, and we henceforth omit the details here.

\begin{mylem}
The generalized Rayleigh quotient  $R_{\mathbf{L_\mathcal{S}},\mathbf{D_\mathcal{S}} }(\mathbf{x})$ is equivalent  to the form of $R_{\mathbf{\bar L_\mathcal{S}}}(\mathbf{D_\mathcal{S}}^{1/2} \mathbf{x})$, where ${\mathbf{\bar L_\mathcal{S}}} = \mathbf{I} - \mathbf{D_\mathcal{S}}^{-1/2}\mathbf{A_\mathcal{S}} \mathbf{D_\mathcal{S}}^{-1/2}$ is the normalized Laplacian matrix of graph $G_{\mathcal{S}}$.  
\end{mylem}
\begin{myproof}
By the definition in Equation \ref{eq:rayleigh} we have
\begin{align*}
R_{\mathbf{\bar L_\mathcal{S}}}(\mathbf{D_\mathcal{S}}^{1/2} \mathbf{x}) &=  \frac{\mathbf{(\mathbf{D_\mathcal{S}}^{1/2} \mathbf{x})}^T\mathbf{\bar L_{\mathcal{S}}}\mathbf{(\mathbf{D_\mathcal{S}}^{1/2} \mathbf{x})}}{\mathbf{(\mathbf{D_\mathcal{S}}^{1/2} \mathbf{x})}^T\mathbf{(\mathbf{D_\mathcal{S}}^{1/2} \mathbf{x})}} \\
&= \frac{\mathbf{x}^T \mathbf{D_\mathcal{S}}^{1/2} \mathbf{\bar L_{\mathcal{S}}} \mathbf{D_\mathcal{S}}^{1/2}  \mathbf{x}}{\mathbf{x}^T\mathbf{D_\mathcal{S}}\mathbf{x}} \\
& = \frac{\mathbf{x}^T(\mathbf{D_\mathcal{S}}-\mathbf{A_\mathcal{S}})\mathbf{x}}{\mathbf{x}^T\mathbf{D_\mathcal{S}}\mathbf{x}} \\
& = R_{\mathbf{L_\mathcal{S}},\mathbf{D_\mathcal{S}} }(\mathbf{x}). 
\end{align*}
\end{myproof}

\begin{mytheo}
(Courant-Fischer Theorem) Let $\mathcal{X}^k$ denote a $k$ dimensional subspace of $\mathbb{R}^N$ and $\mathbf{x} \perp \mathcal{X}^k$ represents that $\mathbf{x} \perp \mathbf{y}$ for all $\mathbf{y} \in \mathcal{X}^k$. For any symmetric matrix $\mathbf{H}\in \mathbb{R}^{N\times N}$ with eigenvalues $\lambda_1^{(\mathbf{H})} \le \lambda_2^{(\mathbf{H})} \le ...\le \lambda_N^{(\mathbf{H})}$, 
\begin{equation}
\lambda_i^{(\mathbf{H})} = \min \limits_{ \mathcal{X}^{N-i-1}} \Big(\max\limits_{\mathbf{x}\perp \mathcal{X}^{N-i-1}, \mathbf{x} \ne 0}  R_{\mathbf{H}}(\mathbf{x}) \Big) =  \max \limits_{ \mathcal{X}^{i}} \Big(\min\limits_{\mathbf{x}\perp \mathcal{X}^{i}, \mathbf{x} \ne 0}  R_{\mathbf{H}}(\mathbf{x}) \Big) 
\end{equation}
\end{mytheo}
We will not include the proof of the Courant-Fischer Theorem here. The interested reader can find a proof in any major linear algebra textbook. 

We denote by $\mathbf{D_\mathcal{S}}^{1/2} \mathbf{x}= \mathbf{z}$. With the Courant-Fischer Theorem, we can express the eigenvalues of the normalized Laplacian matrix $\mathbf{\bar L}_{\mathcal{S}}$ in the following:
\begin{align*}
\lambda_i^{(\mathbf{\bar L}_{\mathcal{S}})} & = \min \limits_{ \mathcal{Z}^{N-i-1}} \Big(\max\limits_{\mathbf{z}\perp \mathcal{Z}^{N-i-1}, \mathbf{z} \ne 0}  R_{\mathbf{\bar L}_\mathcal{S}}(\mathbf{z}) \Big) \\
& = \min \limits_{ \mathcal{Z}^{N-i-1}} \Big(\max\limits_{\mathbf{z}\perp \mathcal{Z}^{N-i-1}, \mathbf{z} \ne 0}   \frac{\mathbf{x}^T(\mathbf{D_\mathcal{S}}-\mathbf{A_\mathcal{S}})\mathbf{x}}{\mathbf{x}^T\mathbf{D_\mathcal{S}}\mathbf{x}}    \Big) \\
&=  \min \limits_{ \mathcal{X}^{N-i-1}} \Big(\max\limits_{\mathbf{x}\perp \mathcal{X}^{N-i-1}, \mathbf{x} \ne 0}   \frac{\mathbf{x}^T(\mathbf{D_\mathcal{S}}-\mathbf{A_\mathcal{S}})\mathbf{x}}{\mathbf{x}^T\mathbf{D_\mathcal{S}}\mathbf{x}}    \Big) \\
& =  \min \limits_{ \mathcal{X}^{N-i-1}} \Big(\max\limits_{\mathbf{x}\perp \mathcal{X}^{N-i-1}, \mathbf{x} \ne 0}   \frac{\sum\limits_{i\sim j} (x_i - x_j)^2}{\sum\limits_i x_i^2 d_i}  \Big) \le 2,
\end{align*} 
where $i\sim j$ indicates that $i$ is adjacent to $j$.
Similarly, 
\begin{equation}
\lambda_i^{(\mathbf{\bar L}_{\mathcal{S}})}  = \max \limits_{ \mathcal{X}^{i}} \Big(\min\limits_{\mathbf{x}\perp \mathcal{X}^{i}, \mathbf{x} \ne 0}   \frac{\sum\limits_{i\sim j} (x_i - x_j)^2}{\sum\limits_i x_i^2 d_i}  \Big) \le 2.
\end{equation}

\begin{mycor}
Given a graph $G_{\mathcal{S}}$ with $N$ nodes, the largest eigenvector of its normalized adjacency matrix is no bigger than 2, i.e., $\lambda_N^{(\mathbf{\bar L_\mathcal{S}})} \le 2$.
\end{mycor}

For any symmetric matrix $\mathbf{H} \in \mathbb{R}^{N\times N}$ with orthonormal eigenvectors $\mathbf{q}_1,\mathbf{q}_2,...,\mathbf{q}_N$, and corresponding eigenvalues $\lambda_1^{(\mathbf{H})} \le \lambda_2^{(\mathbf{H})} \le ...\le \lambda_N^{(\mathbf{H})}$, we can always decompose the binary indicator vector $\mathbf{x}$ into a linear combination of the eigenvectors, i.e., $\mathbf{x}=\sum \limits_i a_i \mathbf{q}_i$. This allows us to write
\begin{equation} \label{eq:rayleigh_weight}
R_{\mathbf{H}}(\mathbf{x}) =\frac{\mathbf{x}^T\mathbf{H}\mathbf{x}}{\mathbf{x}^T\mathbf{x}} = \frac{\Big(\sum \limits_i a_i \mathbf{q}_i^T \Big)\Big(\sum \limits_i a_i \lambda_i^{(\mathbf{H})} \mathbf{q}_i\Big)}{\Big(\sum \limits_i a_i \mathbf{q}_i^T\Big)\Big(\sum \limits_i a_i \mathbf{q}_i\Big)} = \frac{\sum \limits_i a_i^2 \lambda_i^{(\mathbf{H})}}{\sum \limits_i a_i^2} = \sum\limits_i w_i \lambda_i^{(\mathbf{H})},
\end{equation}
where $w_i = a_i^2 / ||\mathbf{x}||^2$. Hence, the Rayleign quotient can be viewed as a weighted average of the eigenvalues. If the indicator vector $\mathbf{x}$ forms an acute angle with the invariant subspace associated with the extreme eigenvalues, then most of the weight in the average must be on eigenvalues close to $\lambda_N^{(\mathbf{H})}$. Similarly, $R_{\mathbf{H}}(\mathbf{x})$ can be bounded from below by the smallest eigenvalues of $\mathbf{H}$, in which case the indicator vector  $\mathbf{x}$ can be approximated by a linear combination of the eigenvectors associated with the smallest eigenvalues.

\begin{mylem}
Let $\mathbf{x}\in \{0,1\}^N$ denote the binary indicator vector for the detected community $\mathcal{C}\subseteq \mathcal{V}_s$ corresponding to the seed set $\mathcal{S}$, the conductance of $\mathcal{C}$ is bounded by
\begin{equation}
\lambda_2 / 2 \le \Phi(\mathcal{C}) \le \min\{1, 2(1-w_1)\},
\end{equation}
where $\lambda_2$ is the second smallest eigenvalue of Laplacian matrix of $G_{\mathcal{S}}$, and $w_1$ is the weight of the smallest eigenvalue of the normalized Laplacian matrix $\mathbf{\bar L_\mathcal{S}}$, as specified in Equation \ref{eq:rayleigh_weight}.
\end{mylem}

\begin{myproof}
The left side inequality holds due to the fact that $\Phi(\mathcal{C}) \ge \phi(G_{\mathcal{S}})$. Following the Cheeger's inequality that $ \phi(G_{\mathcal{S}}) \ge \lambda_2/2$ in Theorem \ref{theo:cheeger}, we therefore have $\lambda_2 / 2 \le \Phi(\mathcal{C}) $. To prove the right side,  we first express the conductance $\Phi(\mathcal{C}) $ using Rayleigh quotient,
\begin{equation}
\Phi(\mathcal{C})  = R_{\mathbf{L_\mathcal{S}},\mathbf{D_\mathcal{S}} }(\mathbf{x}),
\end{equation}
which can be further rewritten as $R_{\mathbf{\bar L_\mathcal{S}}}(\mathbf{D_\mathcal{S}}^{1/2} \mathbf{x})$, according to Lemma 1. Using similar decomposition as that in Equation \ref{eq:rayleigh_weight}, we can express $R_{\mathbf{\bar L_\mathcal{S}}}(\mathbf{D_\mathcal{S}}^{1/2} \mathbf{x})$ as the weighted average of the eigenvalues $\lambda_1^{(\mathbf{\bar L_\mathcal{S}})} \le \lambda_2^{(\mathbf{\bar L_\mathcal{S}})} \le ... \le \lambda_N^{(\mathbf{\bar L_\mathcal{S}})}$, i.e.,
\begin{align*}
R_{\mathbf{\bar L_\mathcal{S}}}(\mathbf{D_\mathcal{S}}^{1/2} \mathbf{x}) &= \sum\limits_i w_i \lambda_i^{(\mathbf{\bar L_\mathcal{S}})} \\
& \le w_1 \lambda_1^{(\mathbf{\bar L_\mathcal{S}})} + (1-w_1)\lambda_N^{(\mathbf{\bar L_\mathcal{S}})} \\
& \le 2(1-w_1).
 \end{align*}
\end{myproof}

\section{Seeding}\label{seed}

Since the initial seed set serves as a key component in our algorithm for uncovering the target community $\mathcal{C}$, it is thus crucial to consider how the quality of seed set affect the performance. In practice, there is not much control over how the seeds are selected. However, the alternative seeding methods can be strategically applied by domain experts in different scenarios based on the availability of candidate seeds. In this section, we will focus on addressing two fundamentally important issues regarding the seed set: 1) What defines ``good" seeds? and 2) How many seeds are needed in order to uniquely define a community?

\subsection{Seeding Method} \label{seeding_method}

To give a well-rounded evaluation on this, we encompass in total five different seeding methods here.  In this experiment, we adopt $|\mathcal{S}|=3$ seeds for each of the seeding method  listed below.
\begin{enumerate}
\item {\bf{High degree seeding}}: pick  $|\mathcal{S}|$ vertices with degree ranked in the top one third among the degree of all vertices in $\mathcal{C}$.
\item {\bf Low degree seeding}: pick $|\mathcal{S}|$ vertices with degree ranked in the bottom one third among the degree of all vertices in $\mathcal{C}$.
\item {\bf Triangle seeding}: pick $|\mathcal{S}|$ vertices in $\mathcal{C}$ that form a triangle as the initial seed set.
\item {\bf Random seeding}:  pick $|\mathcal{S}|$ vertices in $\mathcal{C}$ randomly.
\item {\bf High inward-edge ratio seeding}: the inward-edge ratio for a vertex  $v$ is defined by the fraction of links connecting to another vertex inside the target community $\mathcal{C}$ among all the links coming out from $v$. We pick $|\mathcal{S}|$ vertices with inward-edge ratio ranked in the top one third among all vertices in  $\mathcal{C}$.
\end{enumerate}
\begin{figure*}[t]
\begin{center}$
\begin{array}{cc}
\includegraphics[width=0.85\linewidth]{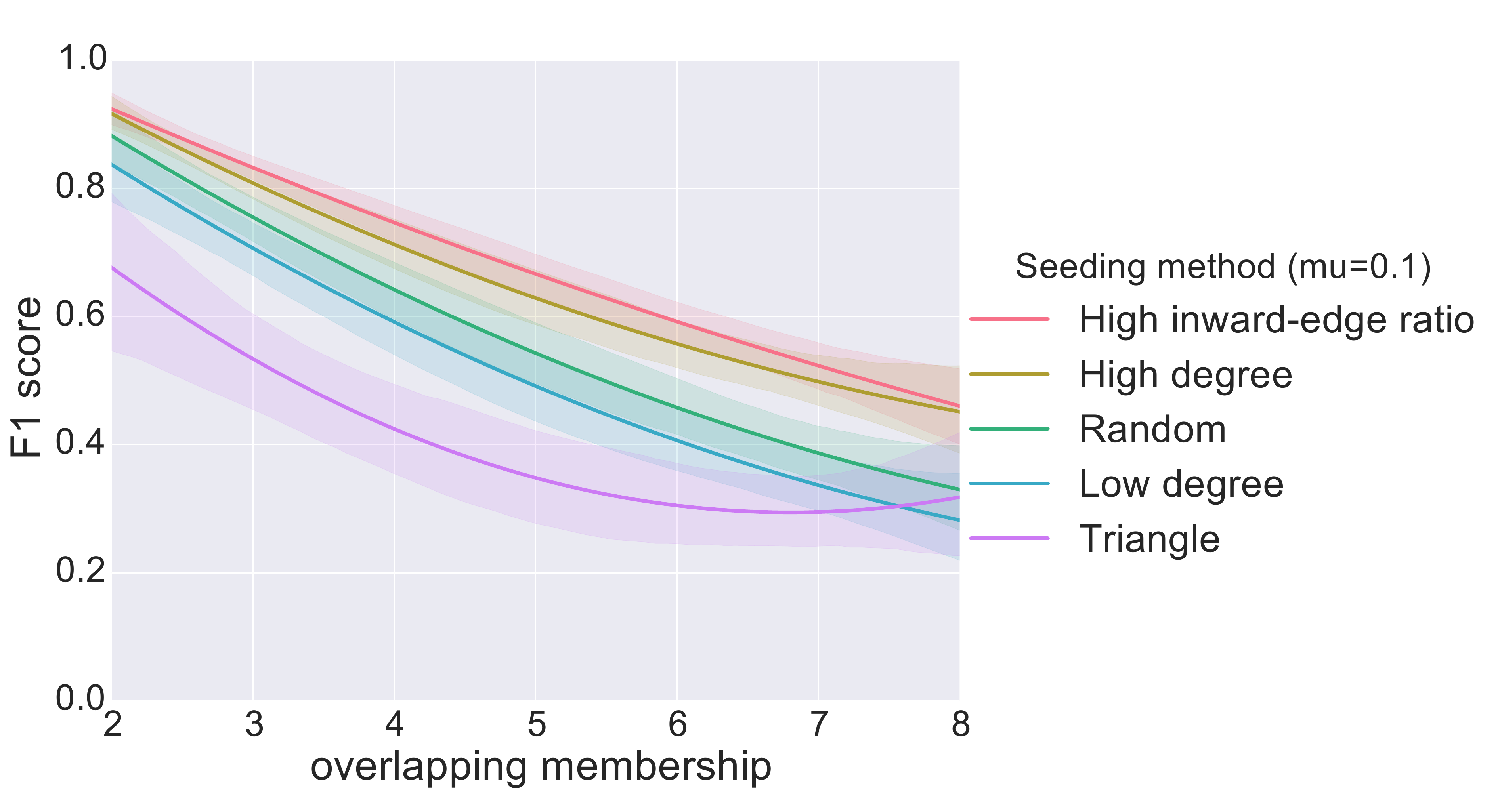}&
\end{array}$
\end{center}
\caption{The average F1 score on LFR datasets ($\mu = 0.1)$ with different seeding methods.  }
\label{LFR_method}
\end{figure*}

\begin{figure*}[t]
\begin{center}$
\begin{array}{cc}
\includegraphics[width=0.85\linewidth]{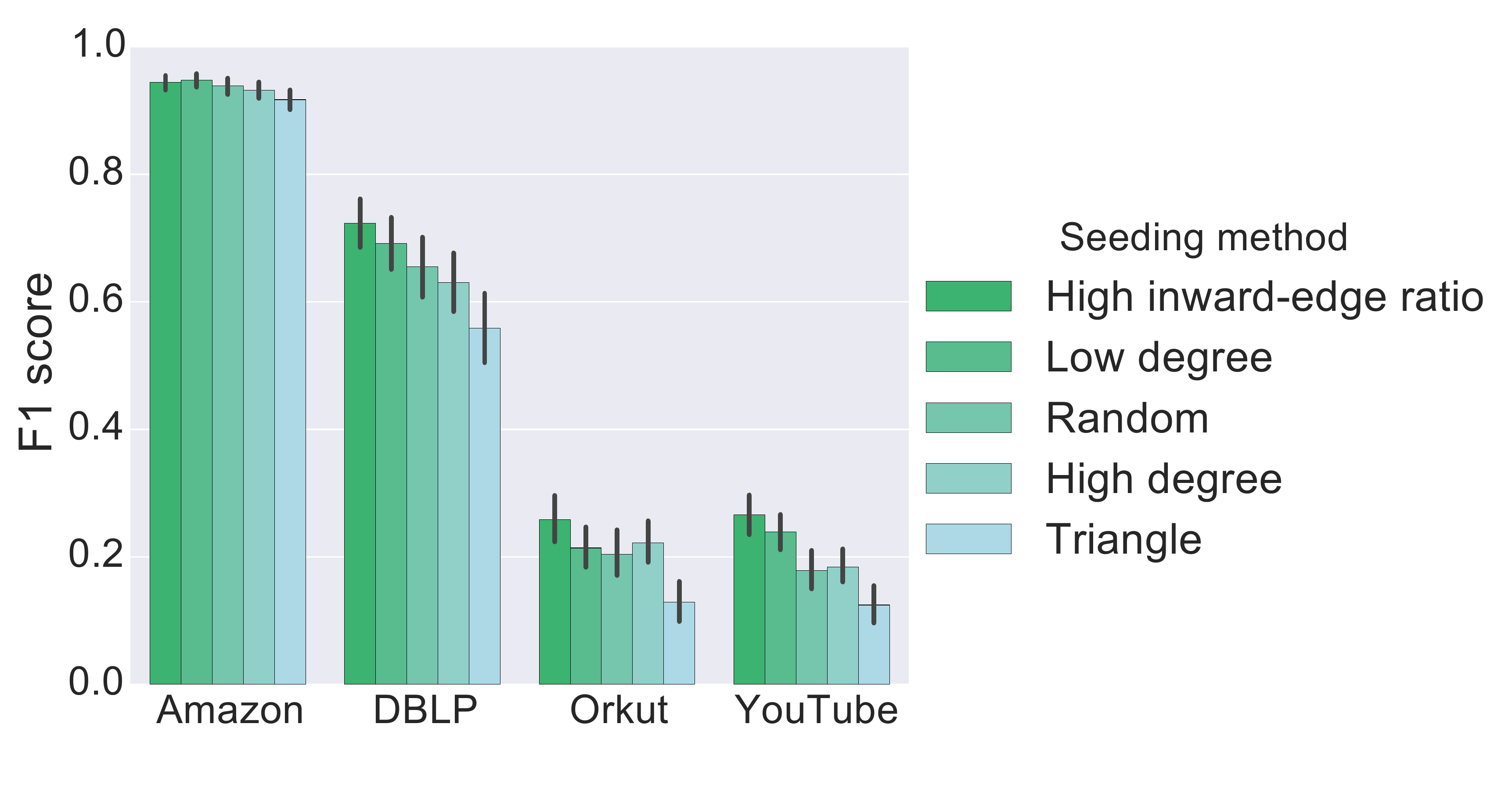}
\end{array}$
\end{center}
\caption{ The average F1 score on real datasets with different seeding methods.  }
\label{real_method}
\end{figure*}

Figure \ref{LFR_method}   gives the experimental results on  LFR benchmark datasets with mixing parameter $\mu=0.1$. It is interesting to note that the high-degree seeding method consistently achieves the highest F1 score in both groups of datasets. When $\mu=0.1$, triangle seeding leads to the worst performance with low F1 score and high standard deviation.  This implies that seeding from a compact core structure is less advantageous than seeding sporadically among vertices. The intuitive explanation behind this phenomenon is that it is more difficult for the probabilities to spread out when the random walk initiates from a cohesive structure. 

Another interesting observation is that high inward-edge ratio seeding method can consistently lead to the best performance among different seeding methods on both synthetic and real datasets. In \cite{Isabel:KDD14}, the authors have the same observation as ours but did not give an explicit explanation on this phenomenon. In fact, when a large fraction of the seeds links connect to vertices within the same community, random walks starting from these seeds would be more likely to transit probabilities into the vertices within the community rather than spreading out to vertices outside the community.  A higher detection accuracy can be thus achieved since the target community contains much of the probability after short random walks.

Moreover, it is also striking to note the difference between the test results on synthetic datasets and that on real datasets. Even though the high-degree seeding method can always bring higher performance than that of random seeding on synthetic datasets, the behavior of these seeding methods on real networks is quite different. In Figure \ref{real_method}, we see that low-degree seeds lead to better result than that of high-degree seeds on DBLP and YouTube datasets. The degree of seeds does not have a significant impact on the performance in Amazon and Orkut networks since the performance of high-degree seeding and low-degree seeding almost tie with each other on these datasets. In \cite{Isabel:KDD14}, the authors compared the detection accuracy  of PageRank based seed set expansion algorithm with high-degree seeding and random seeding on real networks, and concluded that random seeding  method always outperforms high-degree seeding in all domains of real networks. However, we remark here that this observation does not apply to our algorithm as we find that high-degree seeding works slightly better than random seeding on Orkut and YouTube datasets.

\subsection{Seed Set Size}\label{seed_set_size}

It is also interesting to investigate how the size of the seed set affects the performance of our algorithm.

\begin{figure}[htbp]
\begin{center}$
\begin{array}{cc}
\includegraphics[width=0.5\linewidth]{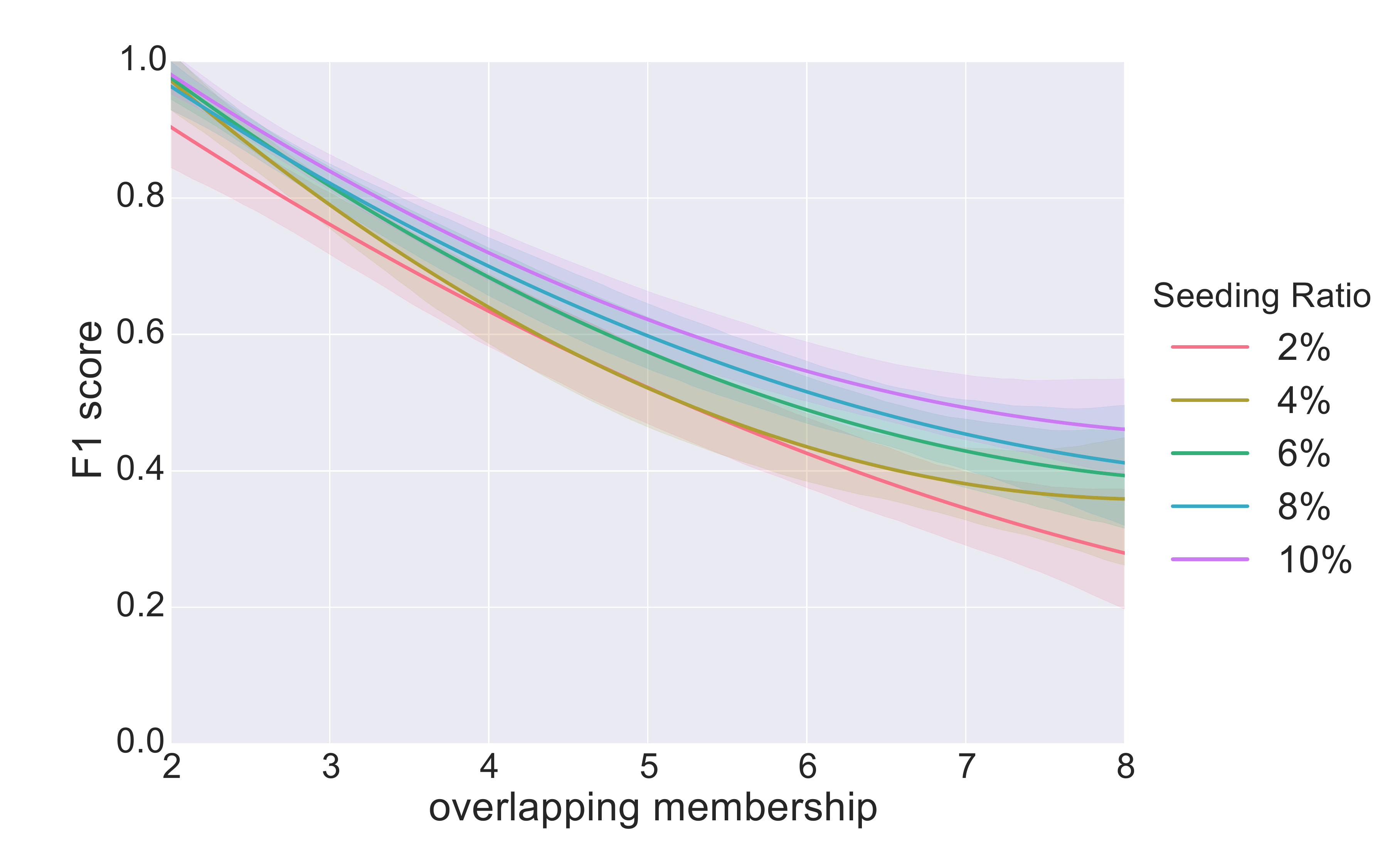}
\includegraphics[width=0.5\linewidth]{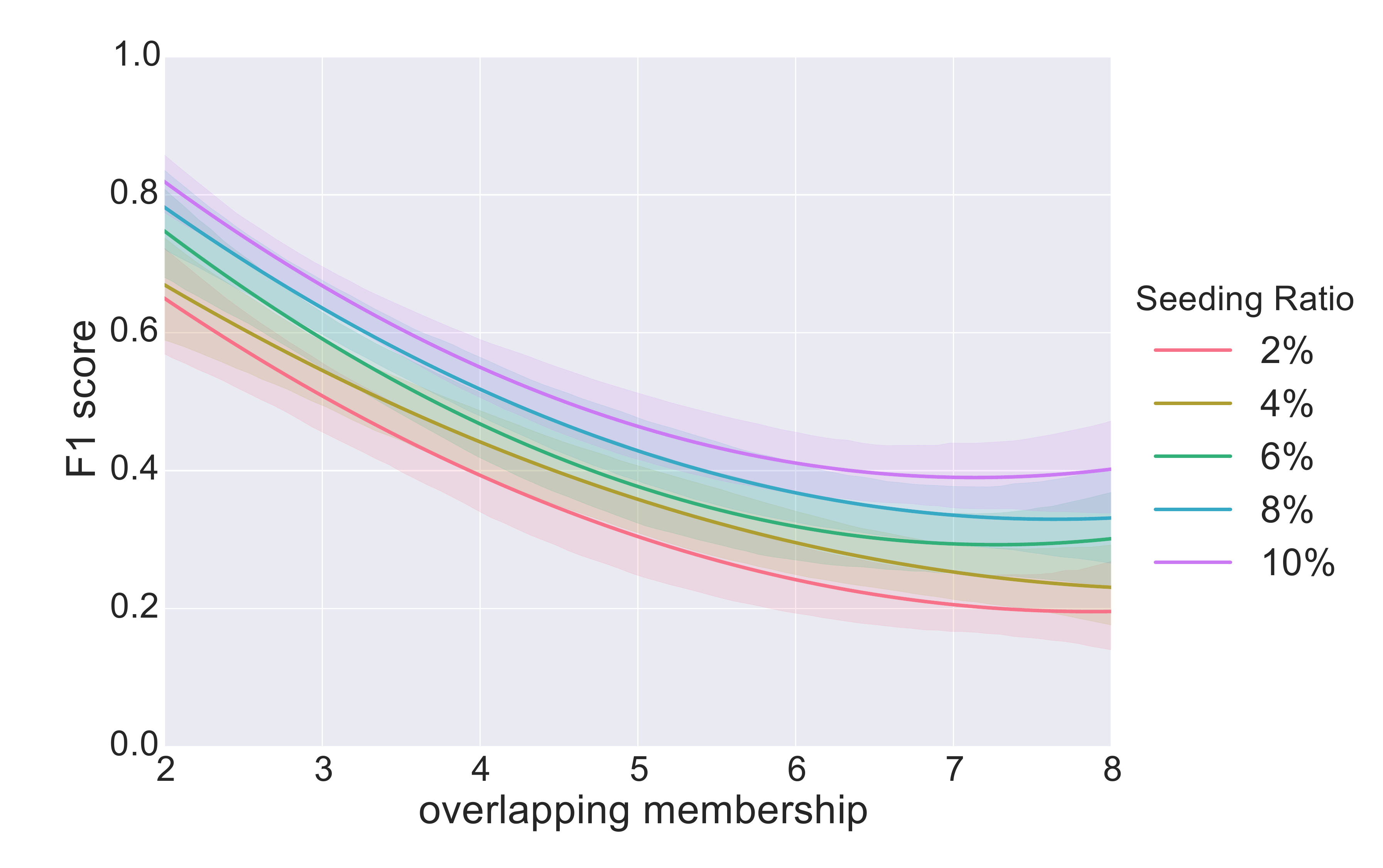}
\end{array}$
\end{center}
\caption{The average F1 score on LFR benchmark data with different seeding ratio. The left figure corre- sponds to the datasets with mixing parameter $\mu$ = 0.1 and the right one corresponds to $\mu$ = 0.3..}
\label{LFR_size}
\end{figure}

We first experiment on the LFR benchmark datasets with varying seed set size. We choose seed set of size proportional to the size of the target community $\mathcal{C}$. Specifically, we test with five different seeding ratios $r$: $2\%$, $4\%$, $6\%$, $8\%$ and $10\%$ respectively, and round $r \cdot |\mathcal{C}|$ to an integer if it is a fraction. Figure \ref{LFR_size} shows the F1 scores when $\mu=0.1$. The algorithm's performance can be improved in general as the seed set size increases. In the case when both mixing parameter and overlapping membership are small, e.g., $\mu=0.1, om=2$,  increasing the seed set size does not seem to affect the performance significantly, and seed set consisting of a small percentage of vertices are sufficient to discover the target community with high accuracy. This implies that when the structure of a small community is well-defined, our algorithm only needs 2 to 3 seeds to reveal the remaining members in a community of size roughly 100. In general, we use an $8\%$ fraction of the vertices in the target community for the whole LFR datasets.

We then carry out the similar experiment on the real datasets.  The result on real networks is interesting because increasing the seed set size has little affect on the performance. Especially, using only 3 seeds can yield almost the same performance as using an $8\%$ fraction of the vertices in the target community as seeds on real datasets.

Our algorithm is thus advantageous to many other seed set expansion algorithms that usually require a higher fraction of vertices to be known. For example, in \cite{Isabel:KDD14}, the authors perform a similar experiment on DBLP network. The performance of their algorithm achieves the maximum recall of $0.3$ when seeding ratio is $10\%$, while \textsc{Lemon} can achieve an average F1 score of $0.66$ with 3 vertices. This makes our algorithm practical  for real networks when it is impossible to collect a large number of seeds.

\subsection{Further Extension } \label{extensions}

As the results of using different seeding methods suggests, high-degree seeds can heuristically lead to better result on synthetic data.  Such heuristic implies that a vertex with higher degree may exert higher impact on shaping the subspace we are looking for, and thus affect the performance by leading to different sparse vectors where we obtain the ``candidates" of the target community from.

In practice, we usually have little control on the seed set. The chance we get a seed set of high-degree members is rare. More often than not, the degree of  seeds is randomly distributed.  We are therefore inspired to tailor our algorithm accordingly in order to emphasize the seeds
with high degree. The modification is rather straightforward: when calculating the initial probability vector $p_0$ to start a random walk from, instead of evenly distributing the amount of probability to each seed, we initialize the probability vector according to the degree of each seed. Formally,
\begin{equation}
p_0(v_i)=\left\{\begin{matrix}
d(v_i)/\text{Vol}(\mathcal{S}) & \text{if~~} v_i\in\mathcal{S} \\
 0 & \text{otherwise}
\end{matrix}\right.
\end{equation}
where $d(v_i)$ denotes the degree of vertex $v_i$. In other words, we enforce a bias towards the high-degree vertices at the beginning of the random walk. Note that each time after the reseeding process, the initial probability vector also needs to be recalculated in the same way.

\begin{figure*}[htbp]
\begin{center}$
\begin{array}{cc}
\includegraphics[width=0.85\linewidth]{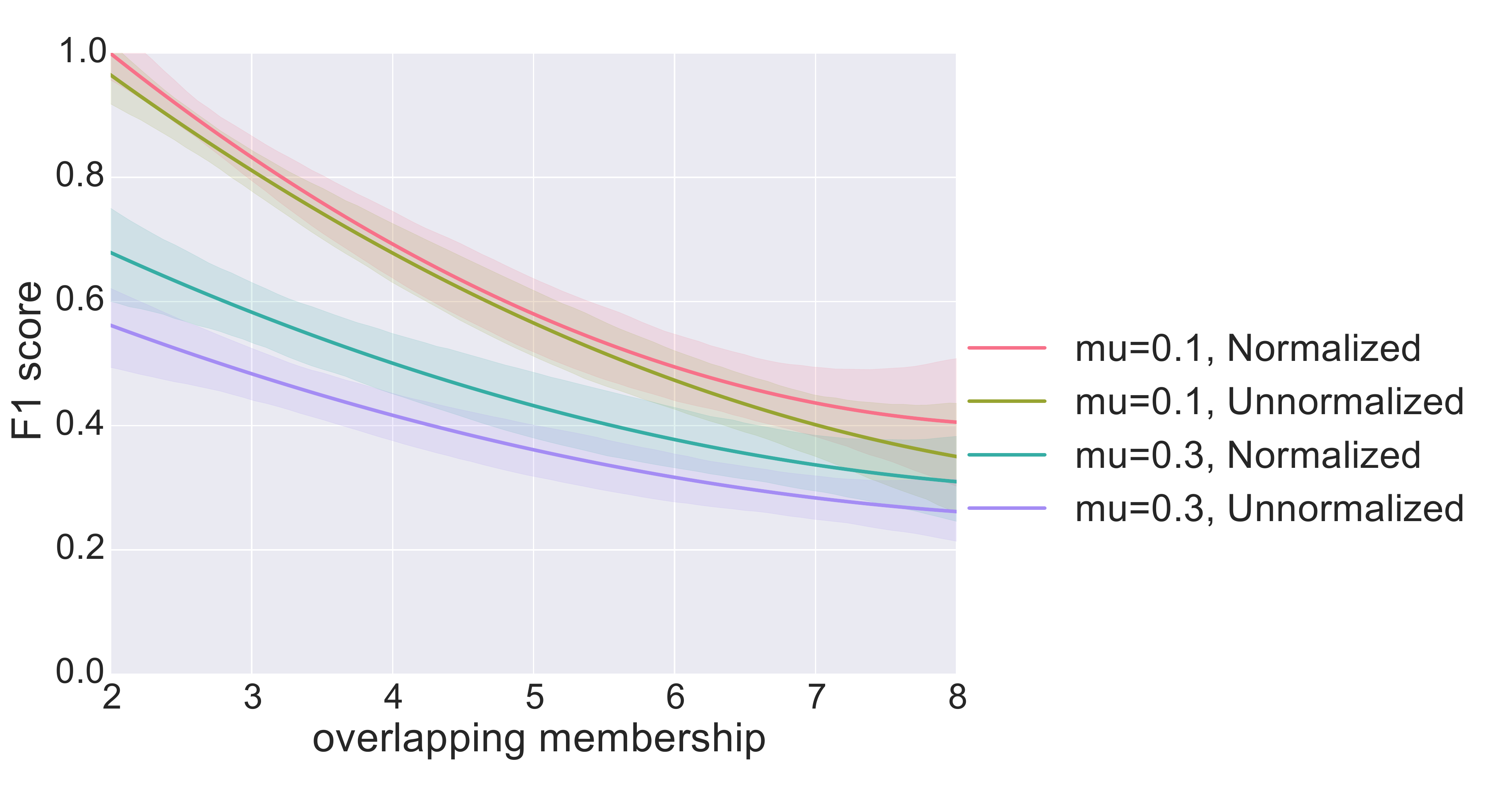} &
\end{array}$
\end{center}
\caption{Comparison of the average F1 score on LFR datasets with and without normalizing the initial probability vector by each seed's degree.}
\label{LFR_bias}
\end{figure*}

Figure \ref{LFR_bias}  depicts the experimental results on LFR benchmark graphs with and without degree normalized initialization for the random walk respectively. We can find that degree-normalization of the initial probability vector results in better performance.

\begin{figure*}[htbp]
\begin{center}$
\begin{array}{cc}
\includegraphics[width=0.85\linewidth]{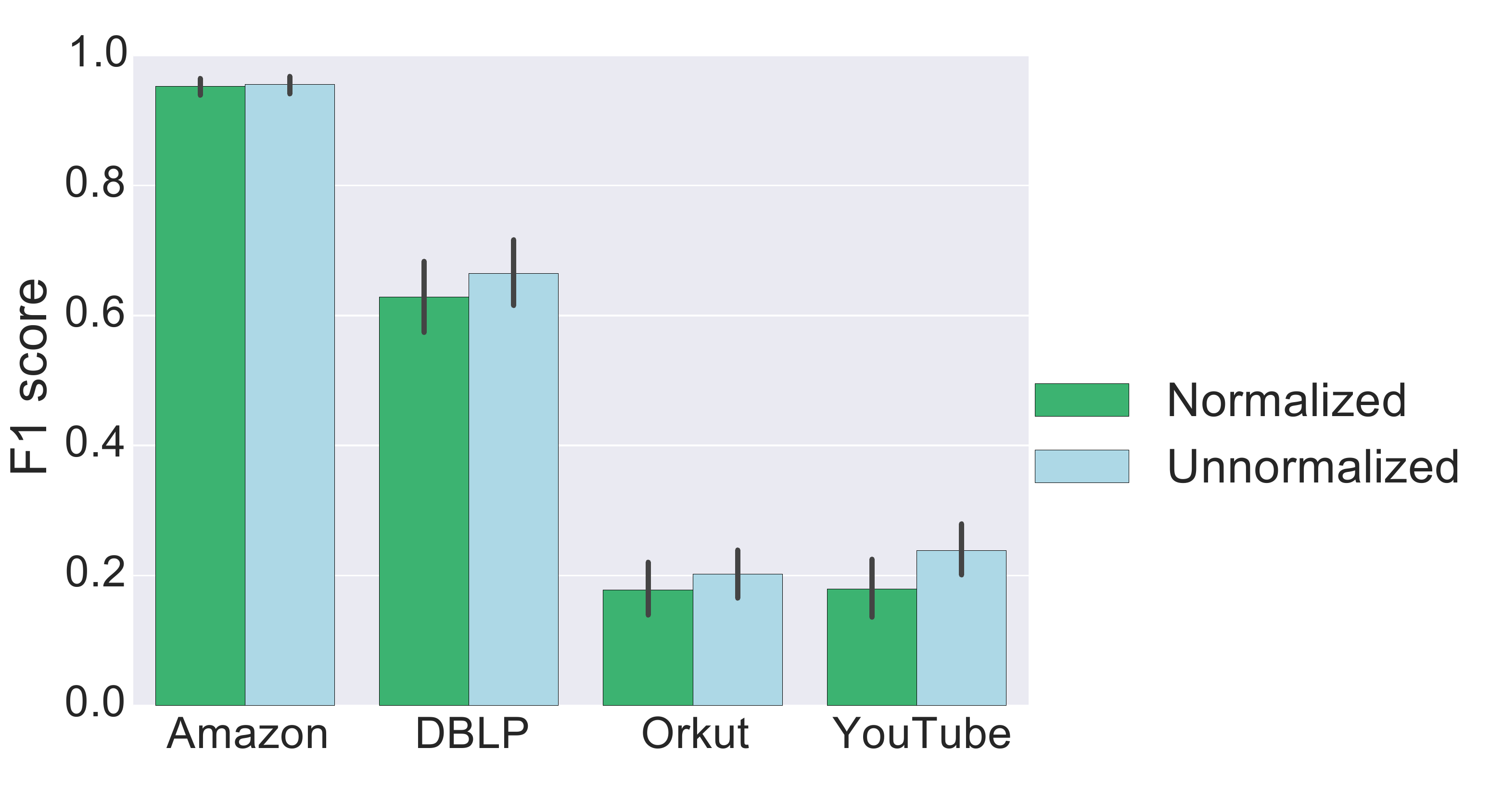}\\
\end{array}$
\end{center}
\caption{ Comparison of the average F1 score on real datasets with and without normalizing the initial probability vector by each seed's degree. }
\label{real_bias}
\end{figure*}

We then perform the same experiments on real networks, and find that degree-normalization would on the contrary, lead to slightly worse statistical results (see Figure \ref{real_bias}). The completely different behavior of using degree normalization on real datasets is rather intriguing.  In fact, this phenomenon accords with our previous observation in Section \ref{seeding_method} that a high-degree seed set is less advantageous than random seeds on real datasets. And this explains why emphasizing on the high-degree vertices would worsen the performance on real datasets.

\subsection{Enlarging the Initial Seed Set}
In Section \ref{seed_set_size}, we see that a larger seed set would lead to better results in general on synthetic datasets. But in the situation when there are not many seeds available, can we still find a way to improve the performance on synthetic datasets? This can be achieved via preprocessing the seed set before running our algorithm.
Specifically, for each pair of vertices $(v_i,v_j)$ in the seed set $\mathcal{S}$, we search for the shortest path $\mathcal{P}$ that connects $v_i$ and $v_j$, and add the vertices on the path to the original seed set if the length of the shortest path  $|\mathcal{P}| \le 3$. The intuition behind this idea is that  any two seeds in the same community must be related for some reason, and they connect with each other either via a direct link or via some other intermediate vertices. In the latter case, those intermediate vertices bridging the seeds are also likely to be in the target community because they serve as the relational ``relay" in order for the seeds to be in the same community.

Note that the procedure of enlarging the initial seed set $\mathcal{S}$ differentiates from the reseeding process while running the algorithm. The pre-processing is done before we feed the seed into the algorithm, which is used for the purpose of increasing the size of initial seed set.

\begin{figure}[htbp]
\begin{center}$
\begin{array}{cc}
\includegraphics[width=0.8\linewidth]{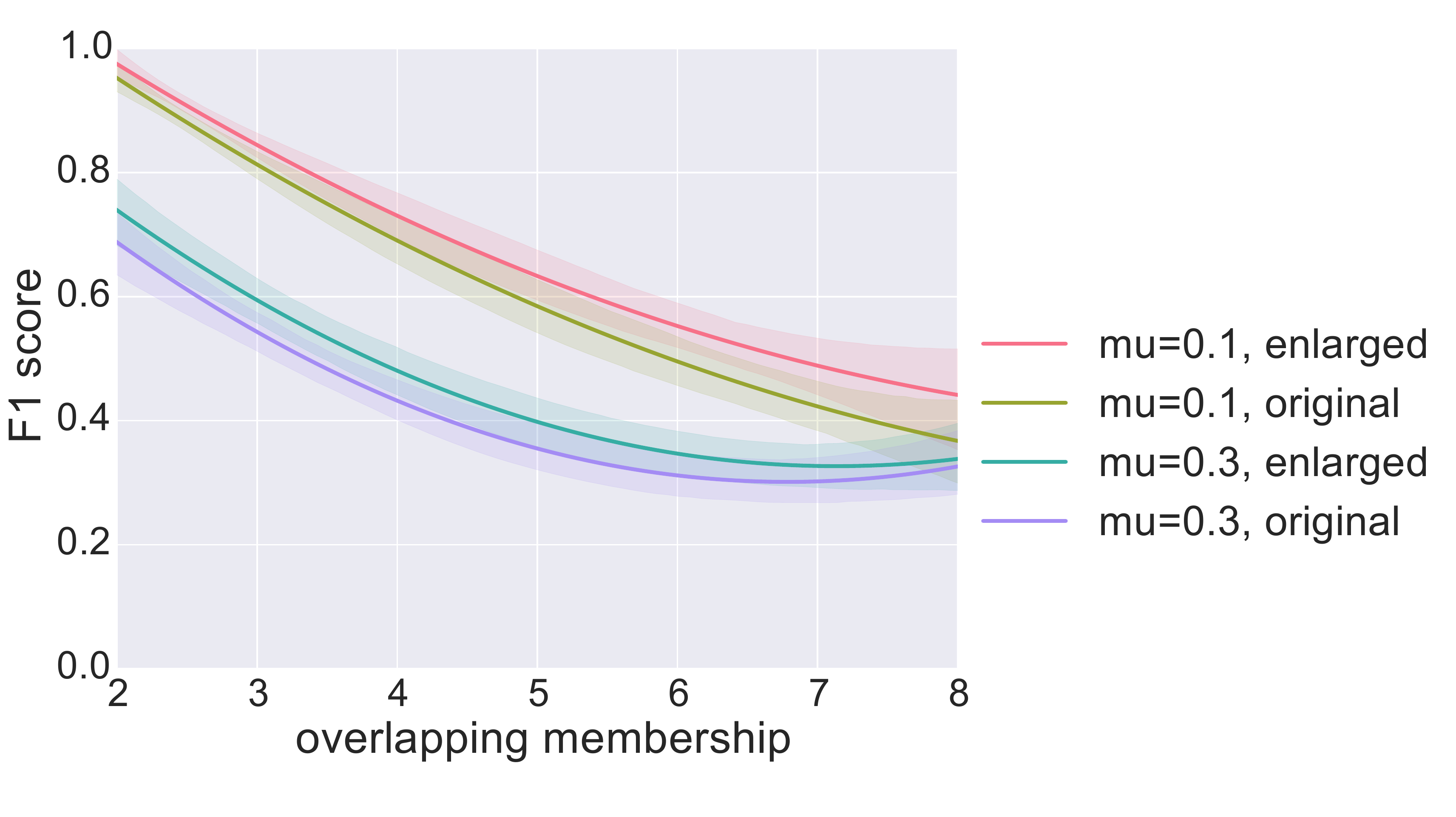} &
\end{array}$
\end{center}
\caption{Comparison of the average F1 score with and without enlarging the initial seed set on LFR benchmark datasets. }
\label{LFR_BT}
\end{figure}

Figure \ref{LFR_BT} presents the experimental results on LFR benchmark data with and without enlarging the initial seed set in advance. We can see that enlarging the seed set can statistically improve the performance. This method can help solve the dilemma of lacking enough available seeds.  Especially, this method would help when the seed set consists of  3 or 4 vertices.

\section{Comparison with the state-of-the-art algorithms}\label{compare}

To give a well-rounded performance comparison with state-of-the-art algorithms, we compared our results to three localized  community detection algorithms and four global community detection algorithms.

\subsection{Comparison with Localized Algorithms} \label{real_evaluation}

\begin{figure}[htbp]
\begin{center}$
\begin{array}{cc}
\includegraphics[width=0.95\linewidth]{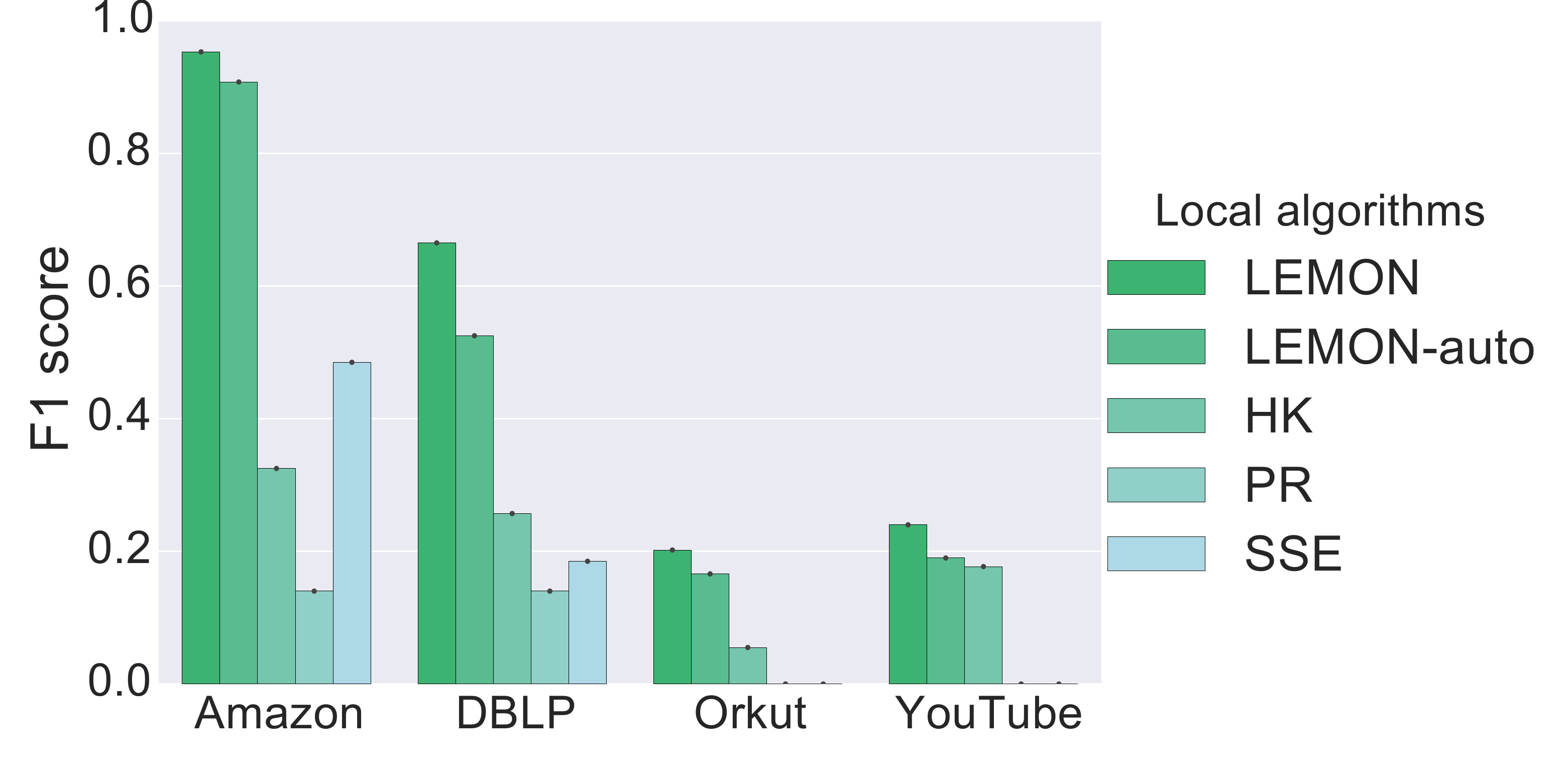} &
\end{array}$
\end{center}
\caption{Comparison of the average F1 score with state-of-the-art local detection algorithms on real networks.  }
\label{real_compare}
\end{figure}

We refer to the experimental results reported in some recent publications on localized community detection algorithms  \cite{kloster2014heat}\cite{Isabel:KDD14}\cite{whang2013overlapping}.  Figure \ref{real_compare} illustrates the comparison of F1 scores on  Amazon, DBLP, YouTube and Orkut datasets. We use ``LEMON-auto" to denote the results obtained by applying the stop criteria in Section \ref{stop}. Since the results on Orkut and YouTube datasets are missing in \cite{whang2013overlapping} and \cite{Isabel:KDD14}, we use empty bars to indicate them.

\begin{enumerate}
\item {\bf Localized algorithms:} We encompass three  locally based methods, Heat Kernel (HK)   \cite{kloster2014heat}, PageRank (PR) \cite{Isabel:KDD14} and Seed Set Expansion (SSE)  \cite{whang2013overlapping}.

\item {\bf Heat Kernel} \cite{kloster2014heat}: The heat kernel \footnote{\url{https://www.cs.purdue.edu/homes/dgleich/codes/hkgrow}}  (HK) is a type of graph diffusion for locally identifying a community nearby a starting seed node. The algorithm can deterministically find the community by computing the diffusion.
\item{\bf PageRank} \cite{Isabel:KDD14}: The personalized PageRank (PR) scheme is computed using the power method and jumpback probability $\alpha=0.10$ in  \cite{Isabel:KDD14}.

\item {\bf Seed Set Expansion} \cite{whang2013overlapping}: The seed set expansion (SSE) approach starts with a phrase of choosing good seed set. The personalized PageRank scheme is then applied to expand the seeds until a community with optimal conductance is found.

\end{enumerate}

Figure \ref{real_compare} show that \textsc{Lemon} achieves an F1 score of 0.910 on the Amazon dataset, far outperforming the other algorithms. The average F1 scores increases the performance  by 3 times compared with the heat kernel algorithm \cite{kloster2014heat} on Amazon, DBLP and Orkut networks. To compare with \cite{whang2013overlapping}, we find that the average F1 score of our algorithm doubles their best performance achieved by the ``spread hubs" method  on Amazon dataset  and triples the performance on the DBLP network. The performance comparison of LEMON and PageRank has been elaborated in Section \ref{compare_pr}.
Also, note that in   \cite{Isabel:KDD14}, the authors did not have an explicit stop criterion and instead assumed using a budget for predicting the size of the target community. We compare with the F1 score at a budget of 100 for both Amazon and DBLP datasets. From the results on Amazon networks in \cite{Isabel:KDD14}, we notice that even granted a budget of 400, which is far beyond the average  community size of 39 in Amazon network, only a recall of 0.45 can be achieved. And we infer the F1 score would be even lower than this value since the precision is dragged down by the large budget set.

It is also worth noting that we only use 3 randomly picked seeds for all the test cases on each dataset. Our algorithm requires very fewer seeds than other algorithms such as \cite{Isabel:KDD14}.

\begin{table*}[htbp]
\begin{center}
\begin{tabular}{l r | r r  r r  r r r r}
\hline
{\bf{Algorithm}}&{\bf{Implementation}}&{\bf{Amazon}} & {\bf{DBLP}} &{\bf{YouTube}} &{\bf{Orkut}}\\
 \hline
\bf{LEMON} &Python &0.953s & 0.665  &0.240 &0.202\\
\bf{LEMON-auto} & Python &0.910  & 0.525  & 0.190  & 0.170 \\
\bf{DEMON} &Python/C++ & 0.164 & 0.196   & 0.031& - \\
\bf{OSLOM}& C++ & 0.766   & 0.542 &  -  & - \\
\bf{LC} &Python/C++ & 0.815& 0.527 & -&- \\
\hline
\end{tabular}
\end{center}
\caption{Comparison of accuracy with global algorithms on real datasets.}
\label{real_accuracy}
\end{table*}%

\begin{table*}[htbp]
\begin{center}
\begin{tabular}{l r | r   r r r}
\hline
{\bf{Algorithm}}&{\bf{Implementation}}&{\bf{Amazon}} & {\bf{DBLP}} &{\bf{YouTube}} &{\bf{Orkut}}\\
 \hline
\bf{LEMON} &Python & $<$15s &  $<$15s &$<$15s &$<$15s\\
\bf{LEMON-auto} & Python  &$<$15s & $<$15s  & $<$15s  & $<$15s\\
\bf{DEMON} &Python/C++ & 4,562s   &727,675s  &  22,395s  & -\\
\bf{OSLOM}& C++ & 885,867s &  23,262s  & $>$10d & -\\
\bf{LC} &Python/C++ & 4,606s &  49,045s &  $>$10d &-\\
\hline
\end{tabular}
\end{center}
\caption{Comparison of running time with global algorithms.}
\label{real_data_time}
\end{table*}%

The experiment has verified that our algorithm is able to achieve high accuracy on large networks constituting communities of average size roughly hundred. This implies that our approach is well-suited for the task of detecting small communities in large networks.

\subsection{Comparison with Global Algorithms}

We also compare local spectral clustering with several state-of-the-art global based algorithms.

\begin{enumerate}

\item {\bf OSLOM} \cite{lancichinetti2011finding}:

OSLOM\footnote{ \url{http://www.oslom.org/software.htm}} is based on the optimization of a fitness function expressing the statistical significance of clusters with respect to random fluctuations (i.e., the random graph generated by the configuration model \cite{molloy1995critical} during community expansion). The worst case running time of OSLOM is $O(n^2)$.

\item {\bf DEMON} \cite{coscia2012demon}:

The DEMON\footnote{\url{http://www.michelecoscia.com/?page_id=42}} algorithm adopts a local-first approach for finding communities. It democratically lets each vertex vote for the communities it sees surrounding it in its limited view of the global system using a label propagation algorithm, and then merges the local communities into a global collection.

\item {\bf LC} \cite{ahn2010link}:

Link Community\footnote{\url{https://github.com/bagrow/linkcomm}} (LC) is a global partitioning algorithm that first builds a hierarchical  link dendrogram according to the link similarity and then cuts the dendrogram at some threshold to yield link communities. The time complexity is $O(nk_{\max}^2)$ where $k_{\max}$ is the maximum vertex degree in the network.
\end{enumerate}

Table \ref{real_accuracy} and \ref{real_data_time} summarize  the average F1 score as well as the running time of each algorithm on real datasets. Among the baselines, OSLOM and LC fail to terminate within 10 days on the YouTube dataset. The OSLOM algorithm can achieve rather good performance but does not scale well. 

In contrast, our algorithm can consistently return the result within few seconds irrespective of how large the entire graph is.

Besides, our algorithm has small memory consumption, and a machine with 4GB RAM can afford to process networks as large as Orkut since the algorithm does not have to store the whole graph in memory.  Moreover, our locally based algorithm is parallelizable because
each seed set expansion can be computed independently. Such property can bring a further performance gain on running time with multi-threaded implementation \cite{whang2013overlapping}.
\begin{figure}[htbp]
\begin{center}$
\begin{array}{cc}
\includegraphics[width=0.7\linewidth]{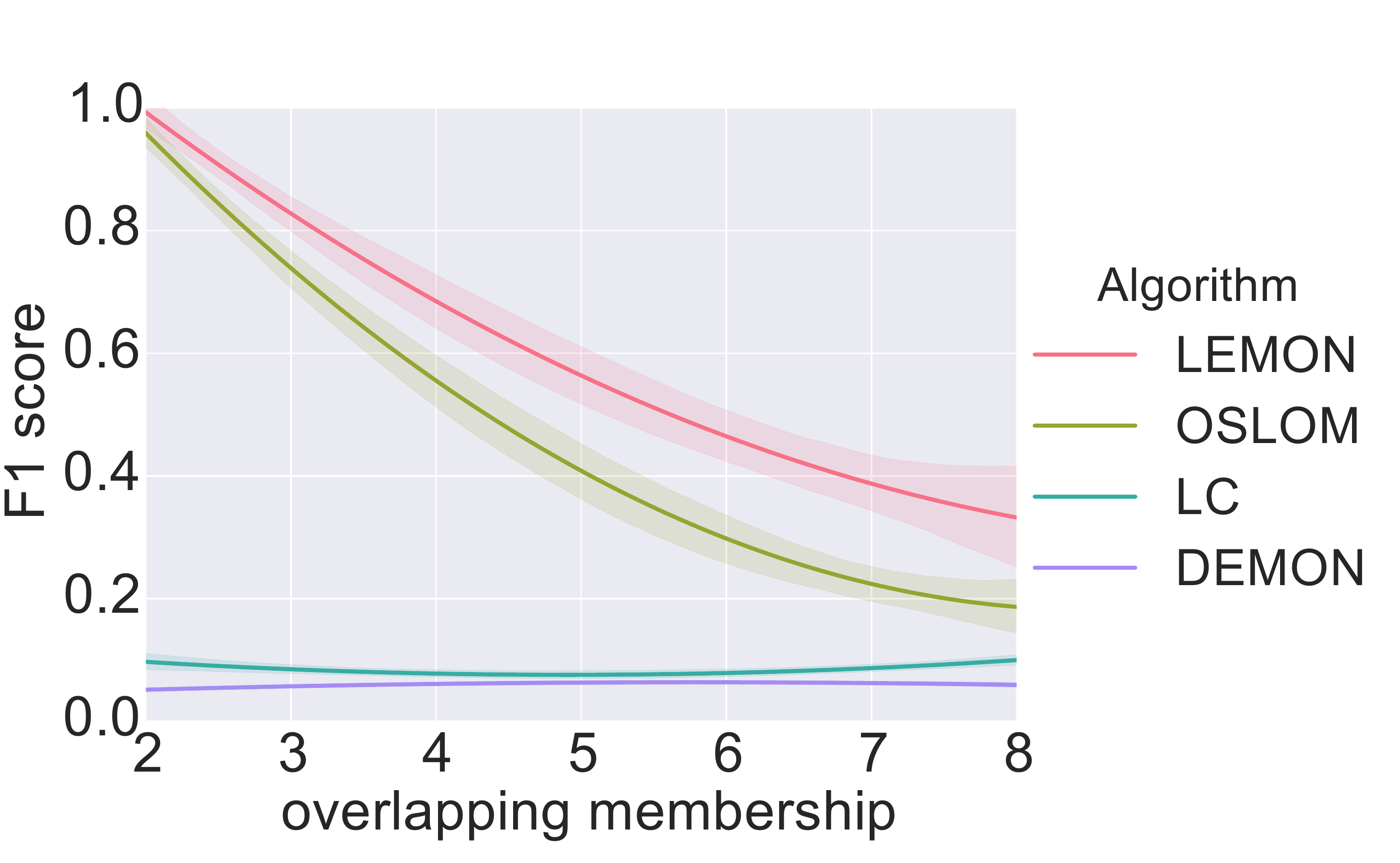} &
\end{array}$
\end{center}
\caption{Comparison of the average F1 score on LFR datasets ($\mu$ = 0.1) with baseline algorithms.}
\label{LFR_compare1}
\end{figure}

\begin{figure}[htbp]
\begin{center}$
\begin{array}{cc}
\includegraphics[width=0.7\linewidth]{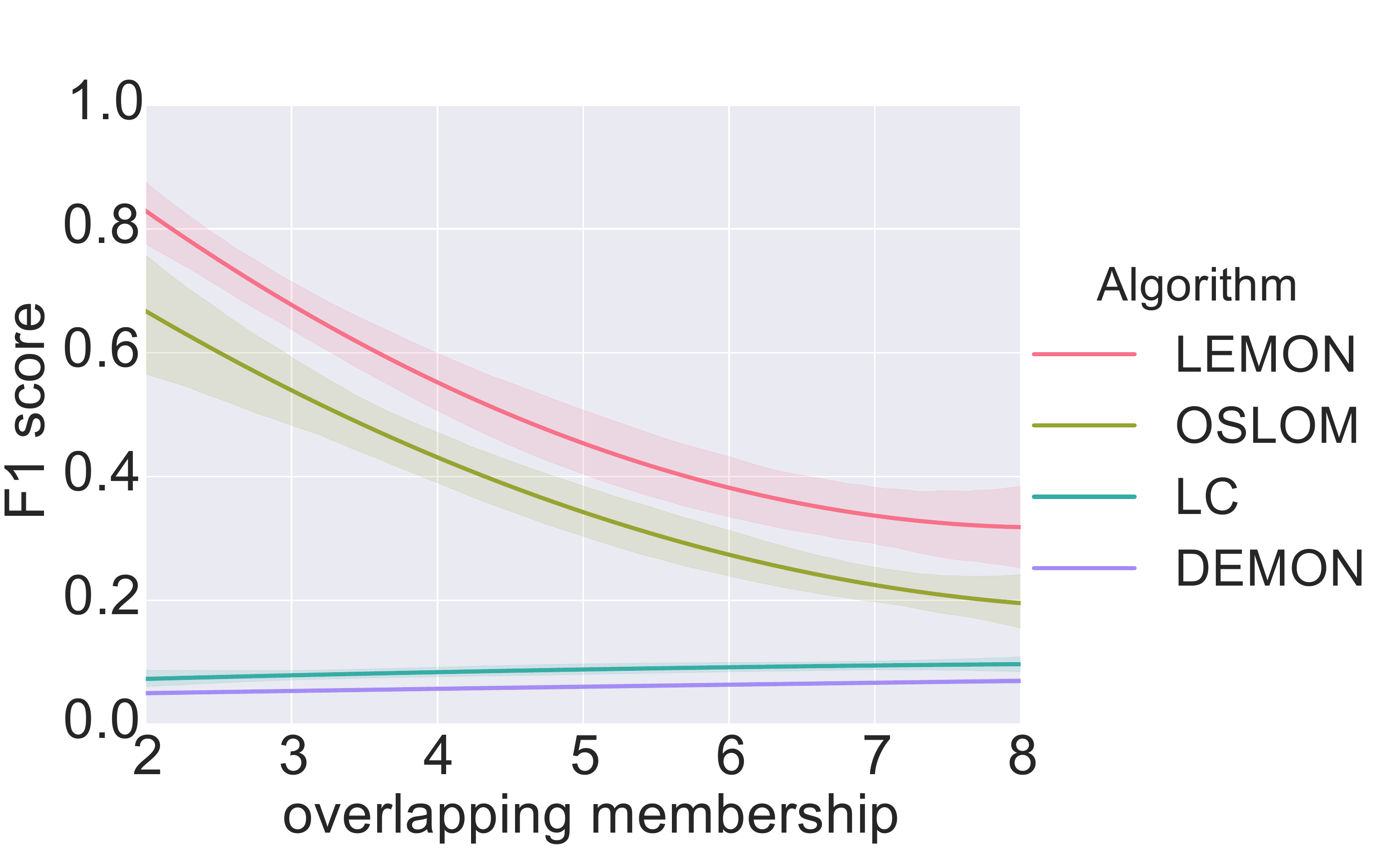} 
\end{array}$
\end{center}
\caption{Comparison of the average F1 score on LFR datasets ($\mu$ = 0.3) with baseline algorithms.}
\label{LFR_compare2}
\end{figure}

 Figure \ref{LFR_compare1} and \ref{LFR_compare2}  compares the average F1 score with some state-of-the-art algorithms on LFR benchmark graphs.  During the experimentation, we also incorporate the methods that can effectively improve the performance on synthetic datasets that are addressed in Section \ref{extensions}. We notice that our algorithm outperforms the baseline algorithms even when we use the random seeding strategy. When the mixing parameter $\mu=0.3$, as is shown in Figure \ref{LFR_compare1} and \ref{LFR_compare2}, \textsc{Lemon} brings about $30\% \sim 40\%$ relative improvement compared with the best results among  the baselines. And we can expect the performance gain to be even more significant if the seeds possess the qualities discussed in Section \ref{seeding_method}.

 Among the four baselines, we notice that LC and DEMON consistently perform poorly on both groups of the synthetic datasets. We further look into the communities found by LC and DEMON respectively, and find that LC tends to partition the graphs into very small pieces while DEMON, on the contrary, usually finds communities that are much larger than the ground truth communities. This implies that both algorithms extract structures from networks that bear little resemblance to the natural formation of the communities. However,  we remark here that even LC fails to recognize the communities well on the synthetic data, it perform better on real datasets as we see in Table \ref{real_data_time} .

\subsection{Empirical Comparison Between Synthetic and Real Data} \label{empirical_comparison}

Networks are not all similar and we cannot assume one algorithm works for finding communities in a network will behave the same on the other networks. Therefore, it is important to develop the understanding of how different types of networks affect the behavior of  algorithms.

Our algorithm sustains a consistent performance on both LFR benchmark graphs and real networks though, we still want to summarize and call the attention to several subtle differences here.

First, \textsc{Lemon} is less sensitive to the parameter of random walk step $k$ and subspace dimension $l$ on real networks than that on LFR benchmark graphs. In practice, fixing $(k,l)$ to be $(3,3)$ for real networks can ensure a good performance. 

Second, \textsc{Lemon} is less sensitive to the seed set size on real networks than that on LFR benchmark. In practice, a seed set size of 3 can guarantee a good performance on  real networks. As for LFR, we adopt the seed set size to be proportional to the community size ($8\%$). 

Third,  \textsc{Lemon} is more sensitive to the high-degree seeds on real networks than that on LFR benchmark. In LFR graphs, the degree of a vertex is at most 50. Whereas in some large real networks such as YouTube, the degree of some vertices exceeds 1000, making the degree distribution much more screw than that seen in LFR graphs. And we expect that vertices with unusually high degree in real networks would have a stronger power in controlling the trend for the probabilities to spread out during the random walk, and thus have a higher risk to enter some other neighboring communities. Such an effect can be counterbalanced by putting less initial probabilities on these ``super cores".

The above empirical analysis informs us that finding communities in real networks seems to be less parameterized than that on synthetic datasets for our algorithm. This indicates that our algorithm is better suited for uncovering those naturally well-formed communities than the artificially constructed communities in practice.

\section{Conclusion}

The problem of identifying small community structure in large networks has been gaining importance. 

In this paper, we have presented a method for finding overlapping communities by seeking a sparse vector in the span of local spectra where the seeds are in its support. To overcome the drawbacks of traditional spectral clustering methods, we propose a novel  method to construct the local spectra  based on the singular vector approximations drawn from short random walks. Our algorithm enables finding a small community in time functional to the size of the community, and it consistently returns the result within seconds even for a network with billions of vertices.  We demonstrate the effectiveness and efficiency of  our method for discovering communities on both synthetic and real-world datasets. As the experimental result shows, our algorithm achieves the highest detection accuracy amongst the state-of-the-art proposals.

Many other fundamentally important research questions remain to be addressed. First, the community detection algorithm based on local spectral clustering could be potentially applied to the membership detection problem, i.e., finding all the communities that an arbitrary vertex belongs to. Second, during the process of seed set expansion, we adopt the first low-conductance community as the target community, which usually yields a high resemblance to the ground truth community. It would also be interesting to look further into some larger low-conductance communities and see if a hierarchical structure exists. In this case, some large social group consisting of several small cliques is likely to be discovered.


\section*{Acknowledgement}
The first author would like to thank Kyle Kloster for the comments on the earlier conference proceeding version. This research has been supported in part by US Army Research Office W911NF-14-1-0477 and National Science Foundation of China 61173180, 61472147.


\bibliographystyle{plain}
\bibliography{acmsmall-sample-bibfile}



\end{document}